\documentclass[%
reprint,
superscriptaddress,
longbibliography,
bibnotes,
 amsmath,amssymb,
 aps,
pre,
floatfix,
]{revtex4-2}

\usepackage{graphicx}
\usepackage{dcolumn}
\usepackage{bm}
\usepackage[normalem]{ulem}
\usepackage{xcolor}

\usepackage{pbox}
\usepackage{tabularx}

\usepackage{appendix}
\usepackage{chngcntr}
\usepackage{etoolbox}
\usepackage{lipsum}
\usepackage{siunitx}

\AtBeginEnvironment{appendices}{%
\addcontentsline{toc}{section}{Appendices}
\counterwithin{figure}{section}
\counterwithin{equation}{section}
\counterwithin{table}{section}
}

\begin{document}

\title{Predicting Defect Stability and Annealing Kinetics in Two-Dimensional PtSe$_2$ Using Steepest Entropy Ascent Quantum Thermodynamics}

\author{Aimen Younis}
\email{younis@vt.edu}
\affiliation{Mechanical Engineering Department, Virginia Tech, Blacksburg, Virginia 24061, USA}

\author{Fazel Baniasadi}
\email{fbasadi@vt.edu}
\altaffiliation[Now at ]{Intel Corporation, Hillsboro, OR.}
\affiliation{Materials Science Engineering Department, Virginia Tech, Blacksburg, Virginia 24061, USA\\}

\author{Michael R.~von Spakovsky}
\email{vonspako@vt.edu}
\affiliation{Center for Energy Systems Research, Mechanical Engineering Department, Virginia Tech, Blacksburg, Virginia 24061, USA}

\author{William T.~Reynolds~Jr.}
\email{Corresponding author: reynolds@vt.edu}
\affiliation{Materials Science Engineering Department, Virginia Tech, Blacksburg, Virginia 24061, USA}


\date{2022-06-08} 

\begin{abstract}
The steepest-entropy-ascent quantum thermodynamic (SEAQT) framework was used to calculate the stability of a collection of point defects in 2D  PtSe$_2$ and predict the kinetics with which defects rearrange during thermal annealing. The framework provides a non-equilibrium, ensemble-based framework with a self-consistent link between mechanics (both quantum and classical) and thermodynamics. It employs an equation of motion derived from the principle of steepest entropy ascent (maximum entropy production) to predict the time evolution of a set of occupation probabilities that define the states of a system undergoing a non-equilibrium process. The system is described by a degenerate energy landscape of eigenvalues, and the entropy is found from the occupation probabilities and the eigenlevel degeneracies. Scanning tunneling microscopy was used to identify the structure and distribution of point defects observed experimentally in a 2D  PtSe$_2$ film. A catalog of observed defects includes six unique point defects (vacancies and anti-site defects on Pt and Se sublattices) and twenty combinations of multiple point defects in close proximity. The defect energies were estimated with density functional theory (DFT), while the degeneracies, or density of states, for the 2D film with all possible combinations or arrangements of cataloged defects was constructed using a non-Markovian Monte-Carlo approach (i.e., the Replica-Exchange-Wang-Landau algorithm \cite{vogel2013generic}) with a q-state Potts model. The energy landscape and associated degeneracies were determined for a 2D PtSe$_2$ film two molecules thick and $30 \times 30$ unit cells in area (total of 5400 atoms). The SEAQT equation of motion was applied to the energy landscape to determine how an arbitrary density and arrangement of the six defect types evolve during annealing. Two annealing processes were modeled: heating from $77$ K ($-196^\circ$C) to $523$ K ($250^\circ$C) and isothermal annealing at $523$ K. The SEAQT framework predicted defect configurations, which were consistent with experimental STM images.

\end{abstract}

\maketitle


\section{Introduction \label{intro_sec:level1} 
}

Transition metal dichalcogenide monolayers are a class of two-dimensional materials that can exhibit a variety of point defects. These defects greatly affect the physical properties. For example, the electrical conductivity and magnetic behavior of two-dimensional PtSe$_2$ \cite{Zhang2017} depend upon the type and concentration of vacancies and anti-site defects. 

A recent study noted that five types of point defects can be found in two-dimensional films of $\mbox{PtSe}_{2}$~ \cite{Zheng2019}. 
These defects are mobile at intermediate temperatures and rearrange into self-organizing patterns~\cite{Zheng2019}. In addition, the self-organizing patterns suggest there are energetic interactions among the defects. Although the ground-state energies of different defects have been calculated with density functional theory (DFT) \cite{Absor2017}, it is desirable to develop an approach that can predict: (a) the stability of defects with temperature, (b) interactions among the defects, and (c) how fast the defects rearrange themselves into patterns over time.

This work applies the steepest-entropy-ascent quantum thermodynamic (SEAQT) framework \cite{2019YvSR1} and its associated equation of motion to predict stable equilibrium configurations of point defects in 2D PtSe$_2$ and explore the kinetics with which initial defect configurations relax to equilibrium.  The SEAQT framework is a non-equilibrium, thermodynamic-ensemble approach that was originally formulated to address a number of physical inconsistencies between quantum mechanics and thermodynamics  \cite{hatsopoulos1976-I,hatsopoulos1976-IIa,hatsopoulos1976-IIb,hatsopoulos1976-III,beretta2005generalPhD}, and it has computational advantages that have led to its application in variety of solid-state problems  \cite{2018YvSR,2019YvSR2,2019YvSR3,2020YvSR,2022MvSR}. 

An essential feature of the framework is an energy landscape that associates every possible microscopic arrangement (or configuration) of a system with a discrete energy eigenlevel. The occupation probability and degeneracy of an energy eigenlevel determines the entropy of the level, and the rate of entropy production as the system evolves toward equilibrium arises from the redistribution of the system's energy among its eigenlevels.  In this contribution, the framework is applied to 2D PtSe$_2$ by constructing an effective degenerate energy landscape using defects whose structure is identified with scanning tunneling microscopy and whose energy eigenlevel spectrum is estimated from DFT. The landscape's degeneracies, or density of states, are found using a non-Markovian Monte-Carlo approach (i.e., the Replica-Exchange-Wang-Landau (REWL) algorithm) with a q-state Potts model. The landscape is then used by the SEAQT equation of motion to find the kinetic path through state space that describes the evolution from an arbitrary initial distribution of defects to stable equilibrium. The energy transitions between levels of the landscape provide the underlying first-principle-basis for the kinetic phenomena present. It is these transitions at each instant of time that are captured by the SEAQT equation of motion based on the principle of steepest entropy ascent (maximum entropy production). In the process, the equation of motion satisfies the first and second laws of thermodynamics and the postulates of quantum mechanics provided the quantum mechanical features of the system have been included in the energy landscape. This permits the equation of motion to predict the time evolution of a set of occupation probabilities that define the states of a system undergoing a non-equilibrium process. These probabilities along with the eigenlevel degeneracies are then used to determine the system energy and entropy at each instant of time. 

To tie the state space evolution to microstructural changes, the density of each of the six types of defects observed experimentally in 2D  PtSe$_2$ films is used as a microstructural descriptor to determine the arithmetic average of the defect configurations predicted by the REWL algorithm for each energy eigenlevel. The occupation probabilities predicted by the equation of motion are then used to determine a weighted average for each defect density at each instant of time from which representative defect configurations or arrangements can be found in a second but limited application of the REWL algorithm.
	
This paper is organized as follows: Section IIA provides a description of the synthesis and imaging of the 2D  PtSe$_2$ film modeled here. This is followed in Section IIB with a description of the thermodynamic model of the system and the DFT calculations used to determine interacting and non-interacting defect formation energies. Section IIC then provides a brief discussion of the energy landscape and the REWL algorithm used to determine the degeneracies for each eigenlevel, while Section IID describes the SEAQT equation of motion used in this application. Results are presented in Section III for two annealing processes that terminate at the same stable equilibrium state consistent with the experimental STM images taken after annealing at $523$ K for 7 hours. The two processes are heating from $77$ K ($-196^\circ$C) to $523$ K ($250^\circ$C) and isothermal annealing at $523$ K.

\section{Methods \label{methods_sec:level1} 
}

\subsection{Synthesis and Imaging \label{experimental_sec:level2}}

\begin{figure*}[h!]
\begin{center}
\includegraphics[width=0.9\textwidth]{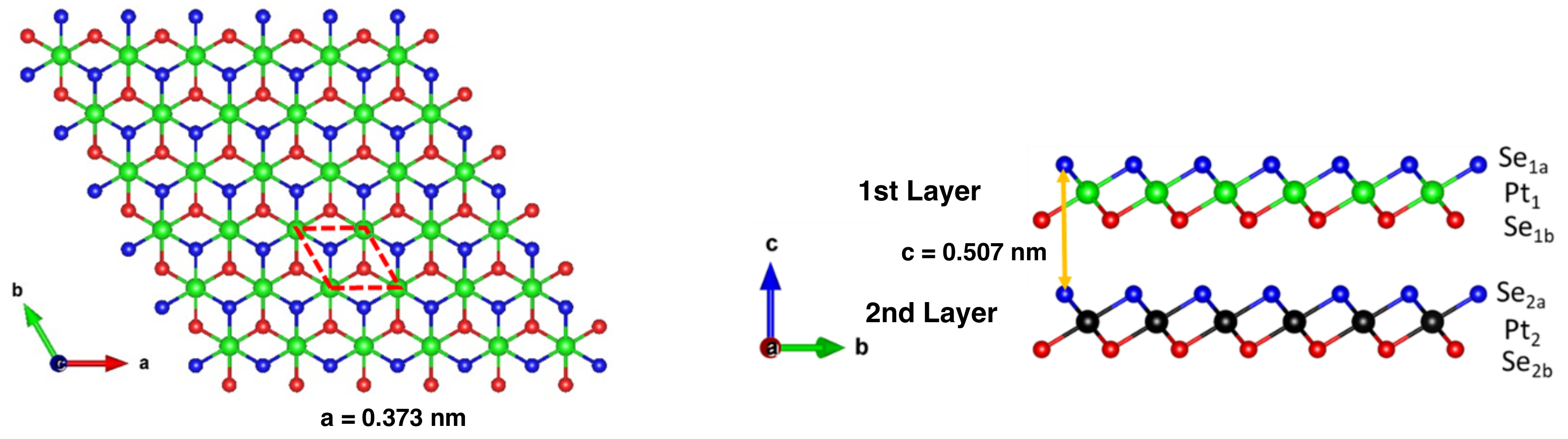}
\caption{Schematic structure of a perfect bilayer of 2D PtSe$_2$. The left image is a projection of the structure along the $c-$axis of the material and the right image is a projection along the $a-$axis.}
\label{fig:1T-PtSe2}
\end{center}
\end{figure*}

\begin{figure*}[h!]
\begin{center}
\includegraphics[width=0.9\textwidth]{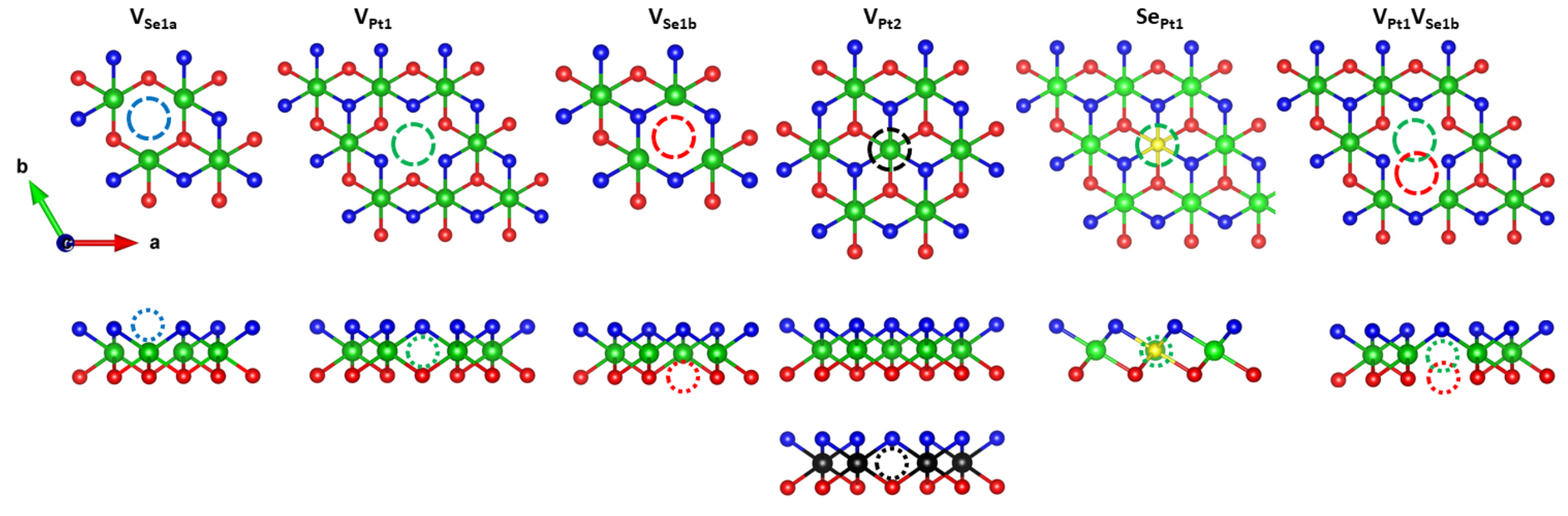}
\caption{Schematic of the six types of point defects found within a unit cell of 2D PtSe$_2$. From left to right: A, B, C, D, E and neighboring B and C defects.}
\label{IndividualDefects}
\end{center}
\end{figure*}

PtSe$_2$ flakes were synthesized via chemical vapor transport on mica. Figure~\ref{fig:1T-PtSe2} shows the structure of a perfect ``monolayer'' of 2D PtSe$_2$, which is represented by two layers of the PtSe$_2$ formula unit. A projection of the unit cell along the $c-$axis is shown to the left in this figure, while a projection along the $a-$axis is shown on the right. As discussed in \cite{Zheng2019}, five different point defects representing either vacancies or anti-site defects were identified in 2D PtSe2, namely, vacancies
$V_{\mbox{\scriptsize Se1a}}$ (A), $V_{\mbox{\scriptsize Pt1}}$ (B), $V_{\mbox{\scriptsize Se1b}}$ (C), $V_{\mbox{\scriptsize Pt2}}$ (D), and anti-site $\mbox{Se}_{\mbox{\scriptsize Pt1}}$ (E) (see schematic representations of these defects in Figure~\ref{IndividualDefects}). Each of these were identified by scanning tunneling microscopy (STM) \cite{Zheng2019}. A sixth ``individual'' defect was defined as the combination of $V_{\mbox{\scriptsize Se1b}}$ and $V_{\mbox{\scriptsize Pt1}}$ located within the same unit cell (again see Figure~\ref{IndividualDefects}). This particular pair of defects was the only combination of defects observed by STM to occur within a single unit cell. Defects were imaged in a $400 \; \mathrm{nm}^2$ ($20\times20 \; \mathrm{nm}$) area using a customized Omicron LT (low-temperature) STM/AFM (atomic force microscope) in constant current mode.

As indicated in H. Zheng $et \; al.$ \cite{Zheng2019}, a STM analysis of and DFT calculations for 1T-PtSe2 were done. Models of the STM images are presented in \cite{Zheng2019} and their agreement with the STM images verified via DFT calculations, which involved integrating the DFT-calculated surface LDOS from the Fermi level to the experimental bias voltage at a plane z = 1 Å above the topmost atomic layer. The results were then compared with the STM images by applying the Tersoff–Hamman approach \cite{Tersoff1983,Tersoff1985}. Integrating the LDOS in this way has been used as a good approximation to STM topographic images in the literature \cite{Kellner2017,Klijn2003,Pan2009} even though isosurfaces of the integrated LDOS relative to constant current are more accurate. VESTA \cite{Momma2011} was then used to visualize the integrated LDOS images. It is this interpretation given in \cite{Zheng2019} regarding the nature of these defects that is used in our present work. 

To do the STM imaging, an initial heat treatment was done on the samples at $250\,^{\circ}\mathrm{C}$ (523 K) for 2.5 hr in an ultra-high vacuum environment (i.e., the preparation chamber of the STM instrument connected to the analysis chamber) with a base pressure of almost 10$^{-10}$ mbar. The samples were then cooled to room temperature for STM analysis. Afterwards, the STM analysis was done at low temperature in order to reduce the thermal motion of the atoms, decrease the thermal and mechanical drift of the sample and STM tip, and improve the temperature stability. Liquid nitrogen was injected around the STM analysis chamber to cool the material to 77 K. After imaging, the sample was heated again to $250\,^{\circ}\mathrm{C}$ (523 K) for an additional 2.5 hr, cooled and analyzed at room temperature and 77 K. This cycle was repeated to achieve cumulative annealing times ranging from 2.5 hr up to 17.5 hr. The analyses at room temperature and 77 K were replicated several times after each heat treatment time. 

\begin{figure*}[h!]
\begin{center}
\includegraphics[width=0.99\textwidth]{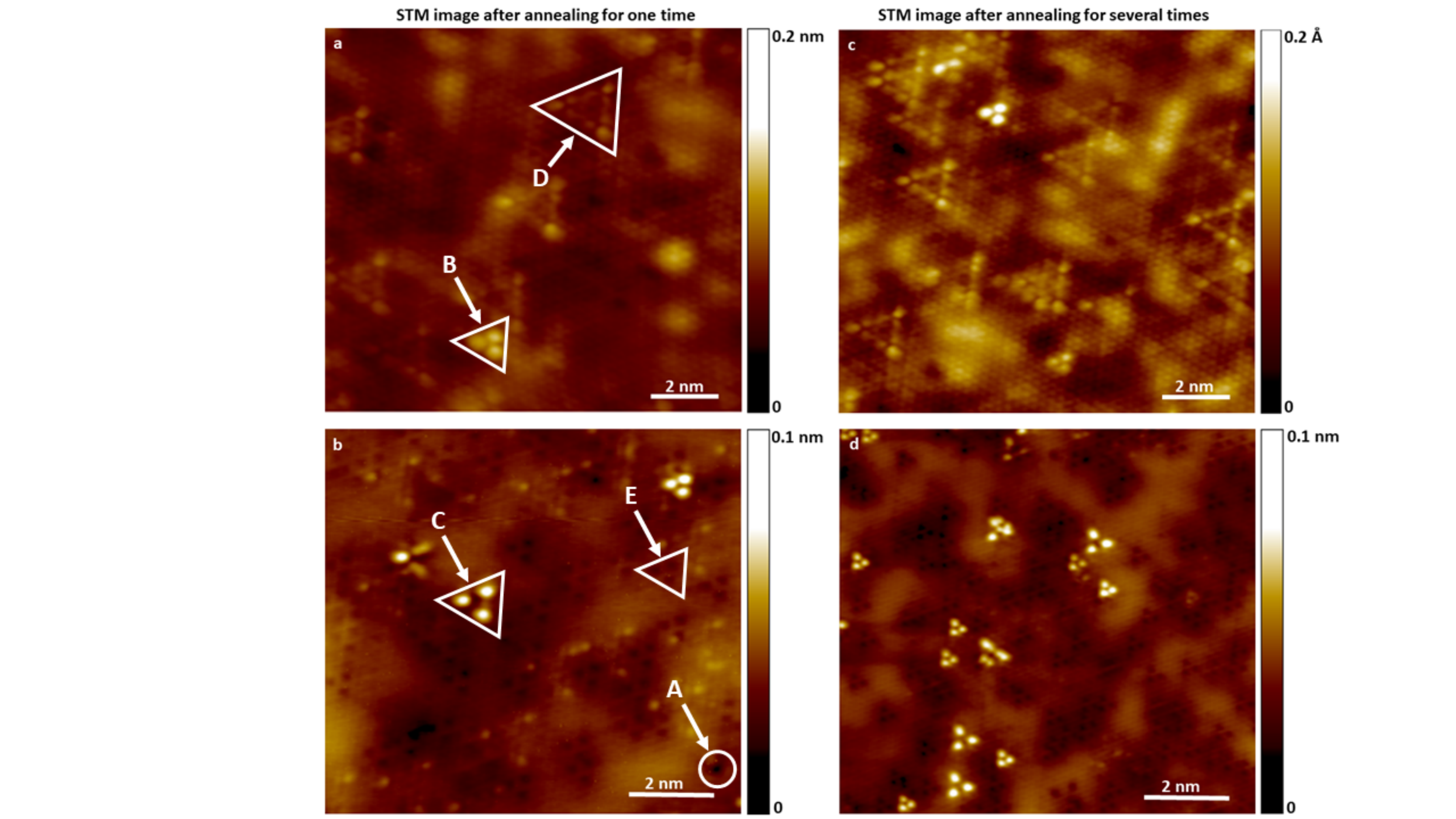}
\caption{STM images of 2D  PtSe$_2$ after annealing at 523 K for 2.5 hr (a and b) and 7.5 hr (c and d).}
\label{DefectMigration}
\end{center}
\end{figure*}

The influence of annealing can be seen in Figure~\ref{DefectMigration}, which shows how the appearance of point defects in a monolayer of 2D PtSe$_2$ changes with annealing time at 523 K. Before explaining how defect configurations change during heat treatment, it should be noted that individual defects in STM images are not single bright/dark dots (which are different from the background). In principle, STM images are contour maps of the surface atoms of the material. Therefore, if an atom from anywhere of the material is missing, it can impact the surface electron density. For instance, for a missing Pt atom in the top molecule of a two molecular layers thick $\mbox{PtSe}_{2}$ film (Defect B), three bright dots are observed in STM images(3 adjacent bright dots in Figure~\ref{DefectMigration}). Actually, as the $3{\mbox{\scriptsize Se1a}}$ atoms have lost their bonding with the missing Pt, those atoms show different electron densities on the surface. Therefore, this configuration in STM images does not show three adjacent defects but a single defect ($V_{\mbox{\scriptsize Pt1}}$). All 5 types of individual defects observed in STM images are marked in  Figure~\ref{DefectMigration} \cite{Zheng2019}.

After measuring the average density of each defect (calculated by averaging over 15 different 20 nm $\times$ 20 nm STM images at each annealing condition) and comparing these densities after the first annealing step (2.5 hr at 523 K) with the corresponding densities after subsequent annealing steps, it was concluded \cite{Zheng2019} that the density of each defect does not change significantly with annealing time {\em at the annealing temperature of} 523 K.  Thus, the evident changes in Figure~\ref{DefectMigration} arise from the migration and rearrangement of defects during annealing. Note that these defects are stable at room temperature, but as the temperature is increased to the annealing temperature (523 K) and the samples kept for a few hours, the defects migrate. Cooling the sample down to the liquid nitrogen temperature (77 K) stabilizes the atoms at their locations after which as outlined in [3] one is able to see the configurational changes in defects via STM.  The SEAQT kinetic predictions presented below show similar changes. Annealing for more hours, as is done in [3], the defects migrate until the migration stops at which point the defect clusters seen in Figure 3 appear. These changes are what motivated the present investigation to predict the equilibrium arrangement and densities of defects as well as the kinetics and densities of the rearrangement process during annealing. In fact, as will be seen below, a comparison of the weighted-average densities predicted by our SEAQT framework to the experimentally derived densities indicate that we quantitatively to the same order of magnitude match what is seen at equilibrium for the densities reported in \cite{Zheng2019}. Furthermore, based on these weighted-average density values, the equilibrium configuration predicted by the SEAQT framework qualitatively matches the rearrangement of single defects and defect clustering observed in the experimental results.

\subsection{Thermodynamic Model \label{landscape_sec:level2}}
A bilayer film with the structure of Figure~\ref{fig:1T-PtSe2} and a $30\,a \times 30\,a$ area (where $a$ is the in-plane lattice constant) is modeled here as a thermodynamic system with discrete energy eigenlevels. The reference or ground state eigen-energy for this system is taken as that of a perfect, stochiometric 2D PtSe$_2$ film and is arbitrarily set to zero. For a film containing populations of several types of point defects in a specific geometric arrangement, higher eigen-energies are found by adding defect energies to the ground state energy, $E_{\mbox{\scriptsize ref}}$. These defect energies include formation energies, $E_\sigma^f$, of non-interacting point defects and the energies resulting from interactions among point defects that are in close proximity to each other. These latter so-called group energies, $E_{\kappa}^{g}$, account for both the formation of the defects and the interactions among them. Summing all of these energies for a given defect film configuration is the energy, $E_j$, for the $j^{th}$ eigenlevel of the 2D film system given by
\begin{eqnarray} \label{EigenEnergy}
E_j &=& E_{\mbox{\scriptsize ref}}\; + E_j^f + E_j^g  \nonumber \\
 &=&  E_{\mbox{\scriptsize ref}}\; + \sum_{\sigma} n_{\scriptstyle \sigma} E_{\sigma}^{ f} 
\;+ \sum_{\kappa} n_{\scriptstyle \kappa} E_{\kappa}^{g}
\end{eqnarray} 
where $n_{\scriptstyle \sigma}$ is the number of non-interacting point defects of type $\sigma$ with the summation taken over all the non-interacting defect types in the film and $n_{\scriptstyle \kappa}$ is the number of different groups of interacting point defects with the summation taken over all the groups in the film. Note that many film configurations can share the same energy,  $E_{j}$, so the energy levels are highly degenerate.

Implementing Equation~(\ref{EigenEnergy}) to generate the discrete energy landscape of all the possible $E_j$ eigenenergies of defective 2D PtSe$_2$ films would require knowledge of the formation energy of all the point defects that could conceivably form as well as the interaction energies associated with the nearly infinite number of different geometrical arrangements of these defects within the film.  This is an intractable problem, but a few simplifying assumptions guided by experimental observations can be used to generate an {\em effective} energy landscape. The first simplification is to include only those structural defects in the energy landscape that are observed in the STM surveys over a total of about $10^5\, \mathrm{nm}^2$ of film area. Other kinds of defects are certainly possible, but if they are {\em not} observed, they are assumed to have formation energies significantly greater than zero, otherwise, they would be seen. This assumption makes it possible to construct an effective energy landscape that is realistic and requires only the formation energies of experimentally observed defects.

The specific point defects observed by Zheng {\em et al.} \cite{Zheng2019} in 2D PtSe$_2$ after annealing in a vacuum at temperatures up to 523 K include: Se vacancies in the two Se planes in the first layer, Pt vacancies in the first layer and second layer, and an anti-site defect corresponding to a Se atom sitting on a Pt site of the first layer.  These five observed point defects \cite{Zheng2019} are listed in Table \ref{defectTypes}. A sixth ``individual'' defect is defined as the combination of $V_{\mbox{\scriptsize Se1b}}$ and $V_{\mbox{\scriptsize Pt1}}$ located within the same unit cell; this defect is labeled ``$V_{\mbox{\scriptsize Se1b}}V_{\mbox{\scriptsize Pt1}}$'' in Figure~\ref{IndividualDefects}. This particular defect pair is the only combination of defects observed with STM to occur within a single unit cell.  The pair is treated as a single defect so as to allow all defects to be located by the unique coordinates of the unit cell in which they reside.  The formation energies of the individual defects are calculated using Density Functional Theory (DFT) as described in Section~\ref{DFT}: ``DFT Calculations.''

The interaction energies among defects are determined by surveying many STM images to identify reproducible patterns. Twenty different patterns were identified. These ``interacting defects'' consist of combinations of between two and five specific point defects in close proximity. The fact that specific patterns are observed repeatedly suggests these particular combinations lower the net energy of the film. This leads to the second simplifying assumption used to construct the energy landscape, namely, that interaction energies can be included in DFT calculations using a supercell large enough to contain a set of defects in the patterns of interest (see also Section~\ref{DFT}). The energy of the supercell's defect arrangement --- which includes the formation energy of the individual defects and their interaction energies --- are then used to determine the right side of Equation~(\ref{EigenEnergy}). 

The maximum interaction distance between defects (defined as the distance beyond which the energy of a pattern of two defects is the same as the sum of the two individual defects) is found from DFT calculations to be about $5\,a$. Defects separated by more than $5a$ are too far apart to interact energetically, and consequently, they can be treated as isolated defects. The energy of each of the twenty unique configurations of interacting defects is calculated from DFT. A larger set of 180 possible configurations is constructed from the twenty arrangements by including all the equivalent symmetry-related configurations.

\subsection{DFT Calculations \label{DFT}}

Calculations of the formation and interaction energies are based on the DFT calculations and experimental work of Zheng $et \; al.$ \cite{Zheng2019} and Baniasadi \cite{baniasadi2021generalPhD}. These DFT calculations used the Projected-Augmented Wave Pseudopotential \cite{blochl1994projector} and included spin-orbit coupling. Relaxed lattice parameters were calculated using two different approximations for the exchange-correlation functional and compared with experimental out-of-plane lattice constants  ~\cite{furuseth1965redetermined,guo1986electronic,kliche1985far}. The Perdew-Burke-Ernzerhof (PBE) generalized-gradient approximation (GGA) \cite{Perdew1996,Perdew1997} overestimated the experimental lattice parameters by 27-29\%, whereas those calculated using the local density approximation (LDA) ~\cite{Hohenberg1964,Kohn1965} were within 2-3\% of the experimental values. The LDA was, thus, chosen because of its better agreement with the experimental lattice constants. Note that hybrid exchange-correlation functionals (e.g., like HSE06), which can represent the electronic properties of the defects more accurately than either GGA or LDA, were not used in this study because of their greater computational burden and the large number of defect configurations that were simulated. As a result, the LDA exchange-correlation functional can be expected to introduce errors in the defect formation energy calculations. Another potential source of error is that the variation in the formation energies of the defects due to the defect charges, which depend on the Fermi level and the synthesis conditions, was not taken into account. Instead, the formation energies are assumed to vary only as a function of the composition. However, this reduced accuracy in the formation energy for each defect does not greatly influence the coarse-grained, overall density of states used by the SEAQT equation of motion to predict the kinetics and configurations of a system of multiple defects (see Section~\ref{REWL_sec:level2}, ``Energy Landscape''). Thus, we believe that this approximation is justified since it is the collective behavior of many defects, which is the primary focus of our study.

The two-layer 2D PtSe$_2$ structure shown in Figure~\ref{fig:1T-PtSe2} is the geometry used for the DFT calculations. Since different defect combinations and spacings are considered, different supercell sizes (Table~\ref{ConfigurationEnergies} in supplementary information) are selected to relax the geometry to find a whole system energy. A vacuum layer of 2.5 nm is added above and below the monolayer to prevent interaction between layers introduced from periodic boundary conditions. The energy cutoff is set to 280 eV and relaxation is done on all supercells until the residual forces are less than 0.001 eV/nm. For supercells larger than $5 \times 5\times 1$, spin-orbit coupling is turned off, and the energy cutoff is reduced to 230 eV. 
 
Now, since the properties of 2D  PtSe$_2$ are sensitive to the number of layers~\cite{Wang2015semicondPtSe2} and vacancies interact with charge~\cite{Gao2017vacancyChargePtSe2,Chang2022catalystPtSe2}, it is important to consider whether these factors are properly incorporated into the DFT calculations of defect formation and interaction energies. In bulk form, PtSe$_2$ is semi-metallic, but, when the number of layers is less than 6, the density of states changes and the material becomes semiconducting. Thus, in~\cite{Zheng2019,baniasadi2021generalPhD}, the density of states of a pristine (defect-free) 6-layer PtSe$_2$ and 2-layer PtSe$_2$ were simulated to confirm the semi-metal to semi-conductor transition, and STM $\frac{dI}{dV}$ curves (i.e., the slope of the tip current-voltage function) were developed for PtSe$_2$ chips consisting of 5 to 9 monolayers. Comparisons between the 6 and 2 monolayer DFT DOS results show that the DOS’s are similar but not the same. This comparison is used by Zheng $et \; al.$ \cite{Zheng2019} and Baniasadi \cite{baniasadi2021generalPhD} to justify the relevance of the 2 monolayer DOS results to their experimental results. Those experimental results clearly show that the STM $\frac{dI}{dV}$ curves are affected by the presence of the different defects. This is consistent with what is shown in \cite{Li2022}, namely, that both the local defect $\frac{dI}{dV}$ and associated DOS are affected by the defect state and charge and as a result will influence the formation and interaction energies. The extent to which this occurs will, of course, depend on the size of the sample material and the density of defects present. Even though the charge of the defects is taken into account in \cite{Li2022} but not in \cite{Zheng2019,baniasadi2021generalPhD}, the formation energies are of the same order of magnitude. More specifically, the formation energies in [34] are higher than the ones used here, and the deviation is larger the greater the formation energy. The defect with the lowest formation energy at the stochiometric composition, defect E (or Se$_{\textrm{Pt1}}$), differs in its value from that in \cite{Li2022} by about 1 eV, while for defect A (or $V_{\textrm{Se1a}}$), the value differs from that in [34] by less than 0.5 eV at stochiometry. The defect with the highest formation energy, defect B (or $V_{\textrm{Pt1}}$), differs in its value by a somewhat larger value of 1.6 eV.  Nevertheless, this larger difference does not affect our results because the large formation energy for defect B makes it a rare one in our simulations, which is consistent with what is seen in the experimental results. Furthermore, as will be seen below, the density of B defects is quite low, and there are only four B defects in any of the time steps of the results presented below.  The E and A defects are the most prevalent ones because they have the lowest formation energies, and in these two cases, the LDS correlation functional used in \cite{Zheng2019,baniasadi2021generalPhD} and the PBE XC functional used in \cite{Li2022} predict similar formation energies. In addition, and importantly, the use of different correlation functionals does not change the order of the defect formation energies.  

As a final note, if we had used the formation energies from the more sophisticated PBE XC functional in our calculations, the sparse defects (B and probably D and F all involve a Pt vacancy) would simply have been somewhat sparser. As to the plentiful defects, A and E, their formation energies using LDA and PBE XC are similar.  Thus, our predicted densities are effectively an upper bound for the sparse defects, and those for the plentiful defects are about the same as would be expected if we had used the PBE XC functional.

Now, the K-mesh for each supercell used in the DFT calculations of \cite{Zheng2019,baniasadi2021generalPhD} is given in Table~\ref{ConfigurationEnergies}. The formation energies of non-interacting point defects  (identifiers $A$ through $F$) and the group energies of interacting point defects (identifiers 1 through 20) are also listed in Table~\ref{ConfigurationEnergies}. These energies are for the synthesis conditions which could be either Se-rich or Pt-rich where ``rich'' indicates that the amount of the element in the synthesis environment is greater than its stoichiometric value. However,  during heat treatment of the material, the Se-rich or Pt-rich synthesis conditions are no longer present, and defects migrate and make new configurations in an essentially stoichiometric film. Therefore, a way is needed to calculate the formation and group energies of non-interacting and interacting defects, respectively, at the stoichiometric ratio of Se to Pt as well as for any concentration in between the Se-rich or Pt-rich conditions, which occurs due to the presence of defects.

To do so, the defect formation and group energies in Equation~(\ref{EigenEnergy}) can be estimated with DFT. Furthermore, the dependence of both $E_\sigma^f$ and $E_\kappa^g$ on concentration can be taken into account with the chemical potentials of Se and Pt such that
\begin{equation} \label{FormationEnergy}
E_{\sigma}^{ f}  = E_{\sigma}  - E_{ \mbox{\scriptsize perfect}} - \left(n_{\mbox{\scriptsize Se}} \; \mu_{\mbox{\scriptsize Se}} + n_{\mbox{\scriptsize Pt}} \; \mu_{\mbox{\scriptsize Pt}}\right) 
\end{equation} and
\begin{equation} \label{GroupEnergy}
E_{\kappa}^{ g}  = E_{\kappa}  - E_{ \mbox{\scriptsize perfect}} - \left(n_{\mbox{\scriptsize Se}} \; \mu_{\mbox{\scriptsize Se}} + n_{\mbox{\scriptsize Pt}} \; \mu_{\mbox{\scriptsize Pt}}\right) 
\end{equation}
Here $E_{\sigma}$ and $E_{ \mbox{\scriptsize perfect}}$ are DFT-calculated energies of $5\times5$ supercells containing a particular type of defect and the perfect (defect-free) supercell, respectively, while $E_{\kappa}$ and $E_{ \mbox{\scriptsize perfect}}$ are DFT-calculated energies of different supercells containing a particular group of defects and the perfect (defect-free) supercell, respectively.  The factors $n_{\mbox{\scriptsize Se}}$ and $n_{\mbox{\scriptsize Pt}}$ are the numbers of Se and Pt atoms added to (or removed from) the film by the creation of a non-interacting defect or a group of interacting defects (see Table \ref{defectTypes}), and $\mu_{\mbox{\scriptsize Se}}$ and $\mu_{\mbox{\scriptsize Pt}}$ are the concentration-dependent chemical potentials for Se and Pt.  The procedure for calculating the chemical potentials is detailed in Appendix \ref{appendix}.

\subsection{Energy Landscape \label{REWL_sec:level2}}

The SEAQT framework is implemented by applying an equation of motion to an energy landscape that represents all the possible energy eigenlevels the system can occupy together with their respective degeneracies \cite{2019YvSR1}. The energy eigenlevels and degeneracies are used to directly calculate the system entropy at each state of the system as it evolves.  

Equation~(\ref{EigenEnergy}) provides a means of calculating the energy eigenvalue of a 2D PtSe$_2$ film for any possible arrangement of the six defect types considered here. To determine the multiplicity (or degeneracy) of each energy eigenlevel, a permutation formula to determine the number of possible permutations the system has could be used.  However, such formulas are impractical for systems larger than a few tens of atoms. For larger systems, the energy eignlevels and degeneracies can be estimated numerically using a non-Markovian Monte Carlo approach such as the Replica Exchange Wang Landau method~\cite{WangLandau2001a,WangLandau2001b,Landau2004}. To do so, a scheme to describe the microstructure of the system is needed. This can be done using an in-plane (two-dimensional) set of pixels. The energy of this microstructure or system can then be found using a $q$-state Potts model. For the system considered here, the integer $q$ (Potts spin) varies from zero (a perfect unit cell) to six ($1 \leq d \leqslant 6$ is for defects types A to F) and captures the configuration of atoms in different unit cells. Each unit cell represents a single node on the lattice and has an associated $q$ number, which reflects whether it is perfect or defective.

The sum of the energies of all the perfect and defective unit cells is a system eigenenergy, $E_j$, of the energy landscape and is given by the two-dimensional 7-state Potts model interaction Hamiltonian, $H_j$, \cite{Braginsky2005,vogel2014scalable,vogel2014exploring,Zhang2019,Hara2015,Bjork2014,Tikare2010}, i.e.,
	\begin{equation} \label{PottsModelH}
	H_j = E_j = E_{\mbox{\scriptsize ref}} + E_j^f + E_j^g
	\end{equation}
where each of the terms to the right of the second equal are equivalent to those provided in Equation~(\ref{EigenEnergy}) but are determined here based on the spin values of the lattice sites of a $ L \times L$ square Potts model lattice of $N$ sites. Taking the energy of a defect-free film as the reference, $E_{\mbox{\scriptsize ref}}=0$ and the rest of the terms to the right of the second equal sign in Equation~(\ref{PottsModelH}) are given by
\begin{equation} \label{PottsModelEfj}
E_j^f = - \underset {n = 1} {\overset {N} {\sum}} \,  \underset {d = 1} {\overset {q} {\sum}} \; D_{d} \, \delta (q_n, d) \, \prod_{z=1}^{Z} \, \delta (q_z, 0) 
\end{equation}
\begin{equation} \label{PottsModelEgj}
E_j^g = - \underset {n = 1} {\overset {N} {\sum}} \, \underset {m = 1} {\overset {M} {\sum}}  \; V_m \, \delta (s_n,m)
\end{equation}
where $\delta$ is the Kronecker delta, which returns a value of $1$, if the two terms of its argument are equal, or $0$, if they are not. 
In Equation~(\ref{PottsModelEfj}), $D_d$ is the formation energy for a non-interacting point defect $d$ (i.e., one of the A to F point defects). In Equation~(\ref{PottsModelEgj}), $V_m$ is the group energy (combined formation and interaction energy) of interacting group defects 1 to 20 in Table~\ref{ConfigurationEnergies} of Appendix \ref{appendix}. Thus, $M=20$ in Equation~ (\ref{PottsModelEgj}). The quantities $D_d$ and $V_m$ are calculated from Eqs.~(\ref{DdGeneral}) and ~(\ref{VmGeneral}) using the parameters calculated with DFT and tabulated in Table \ref{ConfigurationEnergies} of Appendix \ref{appendix}. Since the Potts model itself is in this case limited to 7 spins (i.e., $q_n$ and $q_z$ equal 0 if a given node or site is unoccupied by a point defect or equal 1,...,6 if occupied), its spins are unable to provide sufficient information to account for the number and placement of interacting point defects. To do so, the parameter $s_n$ is introduced, which takes a value between 1 and 20 and accounts for the type, number, and placement of point defects out to the 5$^{th}$ nearest neighbor interacting with the point defect at a given node $n$. It is assumed based on experimental evidence that beyond the 5$^{th}$ nearest neighbor, there is little influence on the point defect at node $n$. Also, note, that if the the group defects found at a given lattice point $n$ do not match group defects 1 to 20 in Table \ref{ConfigurationEnergies} of Appendix \ref{appendix}, $E_j^g$ in Eq, (\ref{PottsModelEgj}) is assigned a large value since this is not a group that is likely to be encountered experimentally. Finally, note that the product in Equation~(\ref{PottsModelEfj}) is used to determine if there are any point defects within the neighborhood of the point defect at node $n$, and if not, all the Kronecker delta values, $\delta(q_z,0)$, are returned as 1. In this product, $Z$ is the number of nearest neighbors out to the fifth nearest.    

\begin{figure*}[h!]
\begin{center}
\includegraphics[width=0.8\textwidth]{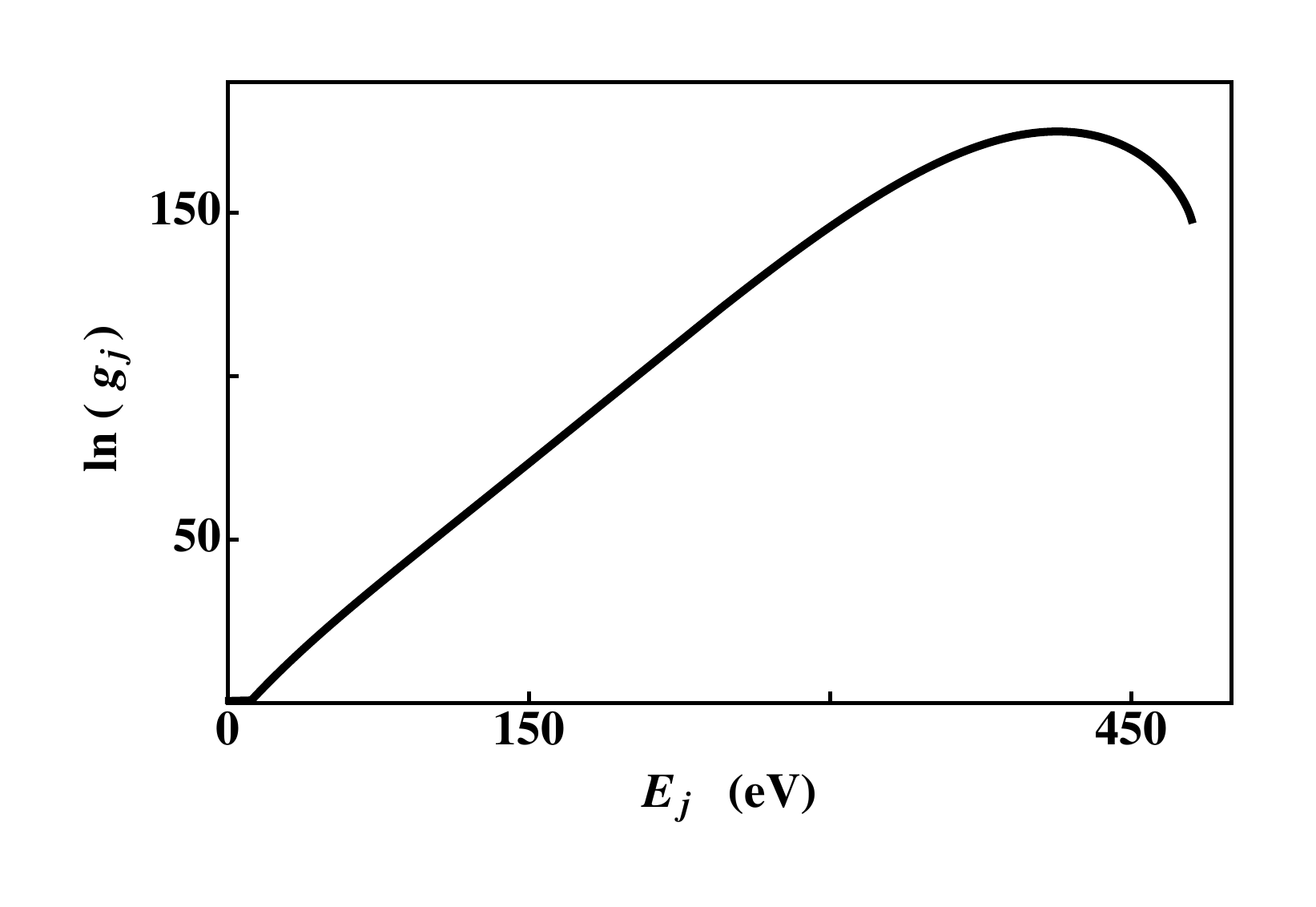}
\caption{Density of states corresponding to the energy landscape of a $30\times30$ 2D PtSe$_2$ lattice containing point defects. The quantity $g_j$ represents the degeneracy of the $j^{th}$ eigenlevel with eigenenergy, $E_j$. The system has 1801 discrete energy eigenlevels: one for a perfect lattice (the lowest energy eigenlevel, $E_0 = 0 \text{ eV}$) and 1800 additional eigenlevels corresponding to different numbers and arrangements of the 6 types of point defects. The highest energy eigenlevel corresponds to $E_j = 480.1 \text{ eV}$. }
\label{fig:DOS}
\end{center}
\end{figure*}

The density of states for the 2D material can be estimated with the Wang-Landau method \cite{WangLandau2001a,WangLandau2001b}, which uses a non-Markovian Monte Carlo walk through all the possible energy levels. The replica exchange \cite{Landau2014protein} variant of the Wang-Landau method greatly accelerates the algorithm by subdividing the energy spectrum into multiple windows, utilizing multiple Monte Carlo walkers over the energy windows and passing information among between the overlapping regions of any two windows.  An implementation of the replica exchange Wang-Landau code~\cite{Vogel2018,vogel2013generic,vogel2014scalable,vogel2014exploring,li2014new} was modified to generate the energy landscape shown in Figure~\ref{fig:DOS}.

\begin{figure*}[h!]
\begin{center}
\includegraphics[width=0.8\textwidth]{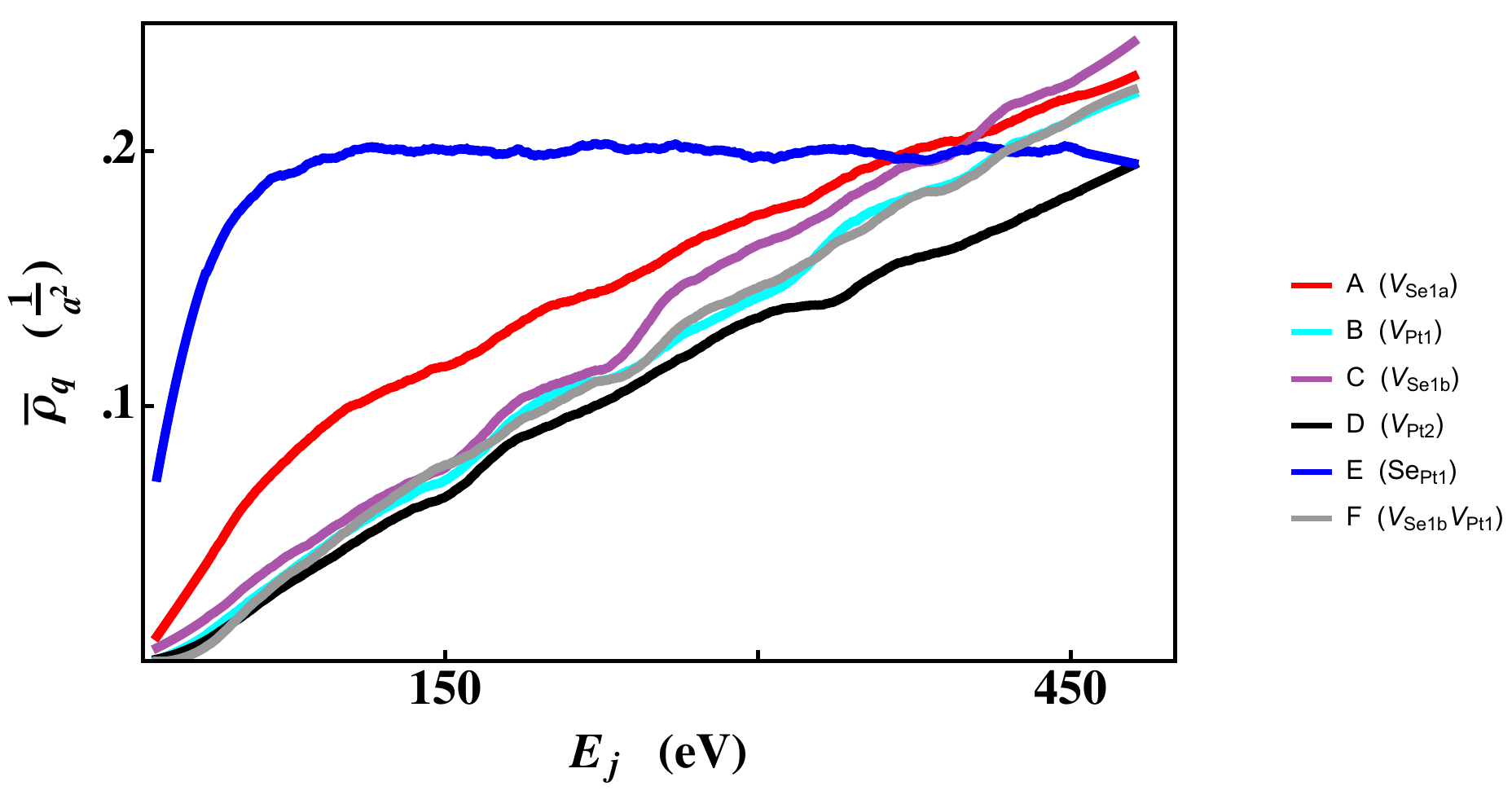}
\caption{Arithmetic-average defect density versus the eigenenergy, $E_j$.}
\label{fig:ArithmeticAve}
\end{center}
\end{figure*}

It is this energy landscape with its energy eigenlevels and associated degeneracies, which is used by the SEAQT equation of motion, described in the next section, to determine a unique kinetic path through state space for a given initial condition. However, in order to connect this path directly to the experimental microstrutures presented here, a procedure developed by McDonald, von Spakovsky, and Reynolds \cite{2022MvSR} is used and entails employing the Replica-Exchange Wang Landau algorithm a second time to count the defects in each configuration sampled along the Monte Carlo walk through the energy eigenlevels. This is done to determine the densities of each defect type, $\bar{\rho}_{\mbox{\scriptsize A}}$, $\bar{\rho}_{\mbox{\scriptsize B}}, \ldots, \bar{\rho}_{\mbox{\scriptsize q}}$, at each eigenlevel as the arithmetic average number of defects over the sampled configurations for a given level. These defect densities are shown as a function of energy eigenlevel in Figure~\ref{fig:ArithmeticAve}. In this figure, the densities of each defect type begins at zero for a perfect, defect-free film and increase smoothly as defects are added to the film and the eigenenergy of the film increases. All the defect densities increase monotonically except the density of defect ``E'', which increases rapidly and saturates. This defect, which is an anti-site defect of Se on a Pt site, has the lowest formation energy of the six defect types considered. These arithmetic densities serve as a set of descriptors that indicate the average number of each type of defect for any specified energy eigenlevel. About 300,000 configurations are used to determine these arithmetic-average defect densities for the 1802 energy eigenlevels. These densities are then used, as described in Section \ref{results:DefectDensities}, to determine the weighted-average densities for each defect at every instant of time using the probability distributions predicted by the SEAQT equation of motion. These latter densities are utilized to construct a typical microstructure at each instant of time and for any expected system energy (see Section \ref{results:Configurations} below).  

\subsection{SEAQT Equation of Motion \label{sec:SEAQT}}

Within the SEAQT framework, the state of a system is specified at every instant of time by a set of energy eigenlevel occupation probabilities, $p_j$. The entropy of each state is calculated directly from the probabilities and their respective degeneracies, $g_j$. The equation of motion used to predict these probabilities in time is based on a geometric construction in state space and the steepest-entropy-ascent principle, resulting in an equation, which inherently satisfies the laws of both quantum mechanics and thermodynamics. The SEAQT equation of motion for any given results in a set of first-order ordinary differential equations in time that can be solved to determine the time dependence of the occupation probabilities from any specified initial state to stable equilibrium.  The kinetics of the system's state evolution are reflected in the time dependence of the occupation probabilities and are a direct result of the energy transition kinetics between energy eigenlevels in time. 

As postulated by Beretta \cite{Beretta2006,Beretta2009,Li2016d}, the SEAQT equation of motion for a simple quantum system is expressed as
\begin{equation} 
\frac {d \hat{\rho}} {dt}=\frac {1} {i\, \hbar}[\hat{\rho},\hat{H}]+\frac {1} {\tau(\hat{\rho})} {\hat D(\hat{\rho})} \label{EOM1}
\end{equation}
where in this expression, $t$ is time, $\hbar$ the modified Planck constant, $\hat{H}$ the Hamiltonian operator, $\hat{D}$ the dissipation operator, $\tau$ a relaxation parameter, and $\hat{\rho}$ the density or so-called ``state'' operator at each instant of time. The $\left[\cdot\right]$ notation on the right represents the Poisson bracket. The term on the left-hand side of the equation and the first term on the right side, the so-called symplectic term, constitute the time-dependent part of the von Neumann form of the Schr\"odinger equation of motion used in quantum mechanics to predict the reversible evolution of pure states (i.e., zero-entropy states). The second term on the right is there to capture evolutions involving the non-zero-entropy states of irreversible processes. 

In the present application, which is purely classical, the energy landscape is constructed without any quantum mechanical information. As a result, the density operator reduces to a classical probability distribution in operator format. Furthermore, for our application, $\hat{\rho}$ is diagonal in the energy eigenvalue basis of the Hamiltonian, and, as a result, commutes with  $\hat{H}$~\cite{Li2016a,Li2016b,Li2018,Beretta2006,Beretta2009,Li2016d}. Thus, the first term on the right of Equation~(\ref{EOM1}) is zero.  

The essence of this equation for a dissipative system lies in $\hat{D}$. This dissipative operator was originally postulated by Beretta \cite{Beretta1984,Beretta1985}, and is based on a constrained-gradient descent in Hilbert space along the direction of steepest entropy ascent at each instant of time that preserves the system energy and occupational probabilities. For an isolated system in which the only generators of the motion are the Hamiltonian and the identity operators, the equation of motion (Equation~(\ref{EOM1}) reduces to \cite{Beretta2006,Beretta2009,Li2016d}
\begin{equation} \label{EOMisolated}
\frac{dp_j}{dt}=\frac {1} {\tau}\frac{\left| 
\begin{array}{ccc}
 -p_j \ln \frac{p_j}{g_j} & p_j & {E}_j\, p_j \\
 \langle S \rangle & 1 & \langle E \rangle \\
 \langle E\,S \rangle & \langle E \rangle & \langle E^2 \rangle \\
\end{array}
\right|}{\left|
\begin{array}{cc}
 1 & \langle E \rangle \\
  \langle E \rangle & \langle E^2 \rangle \\
\end{array}
\right|}
\end{equation} 
Here, the $p_j$ are the occupation probabilities of the $j^{th}$ energy eigenlevel whose energy is $E_j$ and degeneracy $g_j$. In the SEAQT framework, the von Neumann form for entropy of the $j^{th}$ eigenlevel, $S_j = - \ln \frac{p_j}{g_j}$, is used because it satisfies the necessary characteristics for the entropy required by thermodynamics \cite{Gyftopoulos1997} and provides a simple means of directly calculating the entropy of the system in any of its possible states. Additionally, $\langle \cdot \rangle$ represents the expectation value of a property of the system such as the energy, $E$, the entropy, $S$, the energy squared, $E^2$, or the product of the energy and entropy \cite{Beretta2006,Beretta2009,Li2016d}.

The form of the equation of motion given by Equation~(\ref{EOMisolated}) applies to an isolated  system~\cite{Li2016a} and is used for the case of isothermal annealing.  For the case of a film that is annealed during heating or cooling, this equation can be modified to describe an isolated composite system consisting of a film subsystem interacting with a thermal reservoir much larger than the film. For this case, the equation of motion for a subsystem $A$ experiencing a heat interaction with a reservoir $R$ is given by \cite{Li2016b}
\begin{equation}  \label{EOMwithReservoir}
\frac{dp_j}{dt}=\frac {1} {\tau}\frac{\left| 
\begin{array}{cccc}
 -p_j^A \text{ln}\frac{p_j^A}{g_j^A} & p_j^A &0 & {E}_j^A p_j^A \\
 \langle S \rangle^A & 1 & 0 &\langle E \rangle^A \\
 \langle S \rangle^R & 0 & 1 &\langle E \rangle^R \\
 \langle E\,S \rangle & \langle E \rangle^A & \langle E \rangle^R & \langle E^2 \rangle \\
\end{array}
\right|}{\left|
\begin{array}{ccc}
 1 & 0& 
\langle E \rangle^A \\
  0&1&\langle E \rangle^R \\
   \langle E \rangle^A &\langle E \rangle^R &\langle E^2 \rangle\\
\end{array}
\right|}
\end{equation}
Both forms of the equation of motion, Equation~(\ref{EOMisolated}) for isothermal annealing and Equation~(\ref{EOMwithReservoir}) for annealing during heating, can be reduced by expanding the determinants in these expressions (see Refs. \cite{Li2016a,Li2016b,2019YvSR1,2022MvSR} for details). The reduced forms only depend upon the eigenstructure of the film and the temperature of the reservoir, so the superscripts can be dropped to obtain
\begin{align}
\frac {dp_j}{dt^*} & = -p_j \ln \frac{p_j}{g_j} - p_j \frac{C_2}{C_1} + E_j\, p_j \frac{C_3}{C_1} \label{EOMisothermal} \\[3mm]
\frac {dp_j}{dt^*} & = p_j \left[\left(-p_j \ln \frac{p_j}{g_j} - \langle S \, \rangle \right) -  \beta^R \left( E_j - \langle E \, \rangle \right) \right] \label{EOMheating}
\end{align}
where $C_1$, $C_2$, and $C_3$ are cofactors of the first line of the determinant in the numerator of Equation~(\ref{EOMisolated}) and $\beta^R = \frac{1}{k_b T^R}$ in Equation~(\ref{EOMheating}) with $T^R$ the temperature of the reservoir. The variables $t$ and $\tau$ have been replaced by a dimensionless time
defined as $t^*= \int_0^t \frac{1}{\tau(\vec{p}(t'))}dt'$. In this definition, the dissipation parameter, $\tau$, can either be assumed constant or taken to be a function of the time-dependent occupation probabilities $p_j$ represented by the vector $\vec{p}$.  In what follows, it is assumed that $\tau$ is a constant that scales the dimensionless time, $t^*$ to actual annealing time.

Either of these equations form a system of coupled, first-order ordinary differential equations in time -- one equation for each discrete energy eigenlevel of the energy landscape. Solving each system of equations requires an initial condition that correspond to the initial occupation probabilities of all the discrete energy eigenlevels for the system. Solving either system of equations provides the time-dependent occupation probabilities for the energy eigenlevels and, thus, the unique non-equilibrium thermodynamic path taken by the system via its energy landscape.

To obtain the initial condition needed to solve either the system of equations, Equation~(\ref{EOMisolated}) or Equation~(\ref{EOMwithReservoir}), the scheme outlined in reference~\cite{beretta2006steepest} is used here. It utilizes both the canonical and partially canonical distributions $\vec{p}^{\;se}$ and $\vec{p}^{\;pe}$ to determine the initial probability distribution, $\vec{p}^{\;init}$,  needed for the SEAQT equations of motion, using the following perturbation function:
\begin{equation} 
p_j^{init}(t_0) = \lambda \; p_j^{pe}(E_j,\delta) + (1 - \lambda)\; p_j^{se}(E_j) \label{EOMintialDistribution}
\end{equation}
where $\delta$ is the $j^{th}$ value of the vector $\vec{\delta}$ filled with 0's and 1's and with a length corresponding to the total number eigenlevels in the energy landscape. Each 0 indicates no occupation of a given eigenlevel, while 1 means the eigenlevel is occupied. The perturbation parameter $\lambda$ in this equation takes values of $0 < \lambda <1$ and the canonical and partially canonical distributions $\vec{p}^{\;se}$ and $\vec{p}^{\;pe}$ are given by
\begin{equation} 
p_j^{se} = \frac{g_j \; \exp{(- \beta^{se} \, E_j)}}{\sum_j g_j \; \exp{(- \beta^{se} \, E_j)}} \label{EOMcanonicalDistribution}
\end{equation}
\begin{equation} 
p_j^{pe} = \frac{\delta_j \; g_j \; \exp{(- \beta^{pe} \, E_j)}}{\sum_j \delta_j \; g_j \; \exp{(- \beta^{pe} \, E_j)}} \label{EOMpartcanonicalDistribution}
\end{equation}
Here, $g_j$ and $E_j$ are the degeneracy and energy eigenvalue, respectively, of the $j^{th}$ eigenlevel. The values of $\beta^{se}$ and $\beta^{pe}$ are obtained from the constraint on the system energy, i.e., 
\begin{equation} 
\sum_j p_j^{pe}(\beta^{pe})\; E_j = E_{init} \label{EOMbetase}
\end{equation}
\begin{equation} 
\sum_j p_j^{se}(\beta^{se})\; E_j = E_{se} \label{EOMbetape}
\end{equation}
where $E_{init}$ and $E_{se}$ are the energies of the DFT-based configurations at the initial and stable conditions. These energies are determined using the experimental minimum and maximum densities of defects found for each of the point defects in Table 1 of Ref.~\cite{Zheng2019}. For the heating case, the minimum density of defects is used to calculate the total minimum formation energy, $E^{f_{total}}_{min}$, using the point defect formation energies found in Table \ref{ConfigurationEnergies} of Appendix \ref{appendix}. This value is then assigned to $E_{init}$. In a similar fashion, the maximum density of defects from Table 1 of Ref.~\cite{Zheng2019} is used to obtain the total maximum formation energy for the stable equilibrium state (i.e., $E_{se} = E^{f_{total}}_{max}$). For the isothermal case, both the energy for the initial state and that for the stable equilibrium state are equal to each other so that $E_{init} = E_{se} = E^{f_{total}}_{max}$. Once the initial non-equilibrium distribution (i.e., initial state) is found, the SEAQT equations of motion for both cases is solved numerically to predict the state and microstructure evolution in time of the system.

\section{Results and Discussion \label{ResultsDiscussion}}

As noted in Section~\ref{sec:SEAQT}, two different annealing processes were considered: i) isothermal annealing an initial configuration of defects at 523 K  and ii) annealing  during heating from 77 K to 523 K. Because the defect density is not observed to change significantly during isothermal annealing~\cite{Zheng2019}, the initial defect density for this case is matched to experimental AFM images. On the other hand, energy must be added during the second annealing process to heat the film from 77 K to the final equilibrium temperature of 523 K so for this case, a lower initial density of defects with a lower initial energy is assumed, and the defect density is allowed to increase during heating as the energy rises to the level of the final equilibrium temperature. 

\subsection{Kinetic Paths \label{results:paths}}
\begin{figure*}[h!]
\begin{center}
\includegraphics[width=0.8\textwidth]{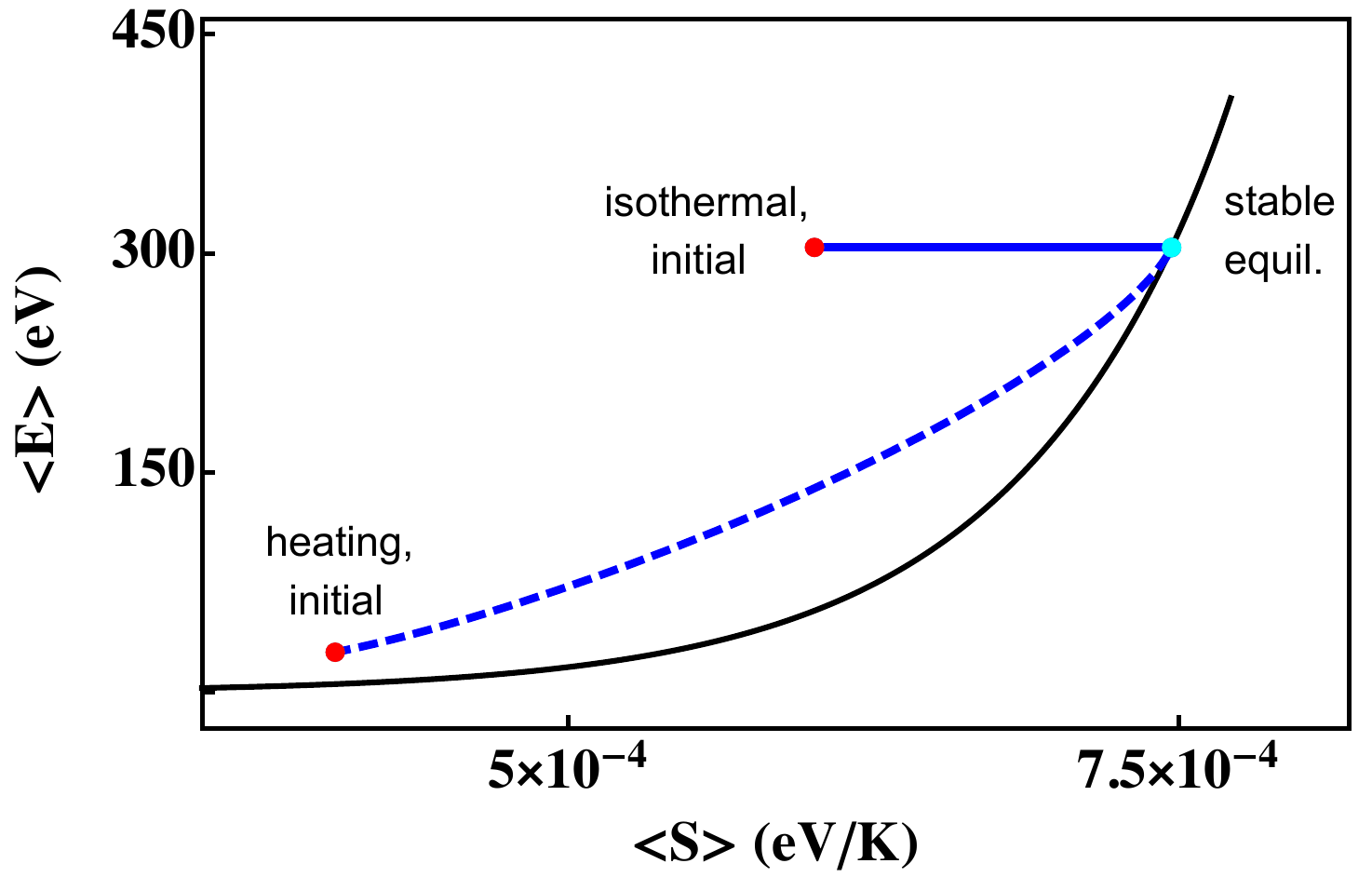}
\caption{Energy as a function of the entropy for the two annealing processes. The solid blue curve represents isothermal (constant $E$) annealing at 523 K. The dashed blue curve represents heating from 77 K to 523 K. The black curve is the set of stable equilibrium states (temperature is defined as the slope of this curve). The red points locate the two different initial states for the annealing processes, and the cyan point is the common final stable equilibrium state.}
\label{fig:EvS}
\end{center}
\end{figure*}

The unique kinetic paths associated with these two processes are determined by the steepest-entropy-ascent principle and are calculated from the time-dependent occupation probabilities that solve the appropriate equation of motion (either Equation~(\ref{EOMisothermal}) or Equation~(\ref{EOMheating})). The two kinetic paths are shown by the solid and dashed blue curves in the energy versus entropy diagram of Figure~\ref{fig:EvS}. The isothermal anneal is a constant energy process represented by the solid blue line in the figure.  The initial states for both processes are represented by the red circles and the final state by the cyan circle. The dashed blue curve represents the path the system follows during annealing with heating from its initial state to its final state, which is the same as that for the first process. The occupation probabilities for the initial states for both processes correspond to perturbed partial equilibrium states constructed using the method described in Section \ref{sec:SEAQT}. 

The final equilibrium states for both annealing processes are identical because both processes terminate at the same energy and temperature. This final state for both processes is predicted by the equation of motion but as well by the canonical probability density distribution of statistical and quantum thermodynamics. Note that the starting and finishing temperatures in the figure are extracted from the experimental annealing conditions rather than the energy landscape since for simplicity only defect energies are included in the landscape~\footnote{The Hamiltonian used to construct the energy landscape (Equation~(\ref{PottsModelH})) only includes defect energies, and, thus, these energies are the only means here for storing energy in the solid. However, a real solid includes phonon vibrational and electron energy modes, and these account for most of the energy of a solid. They are excluded here for the sake of simplicity.  For this reason, tangents to the equilibrium $E$ versus $S$ curve in Figure~\ref{fig:EvS} represent fictitious temperatures rather than actual temperatures.}.

\subsection{Energy level occupation \label{results:probabilities}}

\begin{figure*}[h!]
\begin{center}
\includegraphics[width=0.8\textwidth]{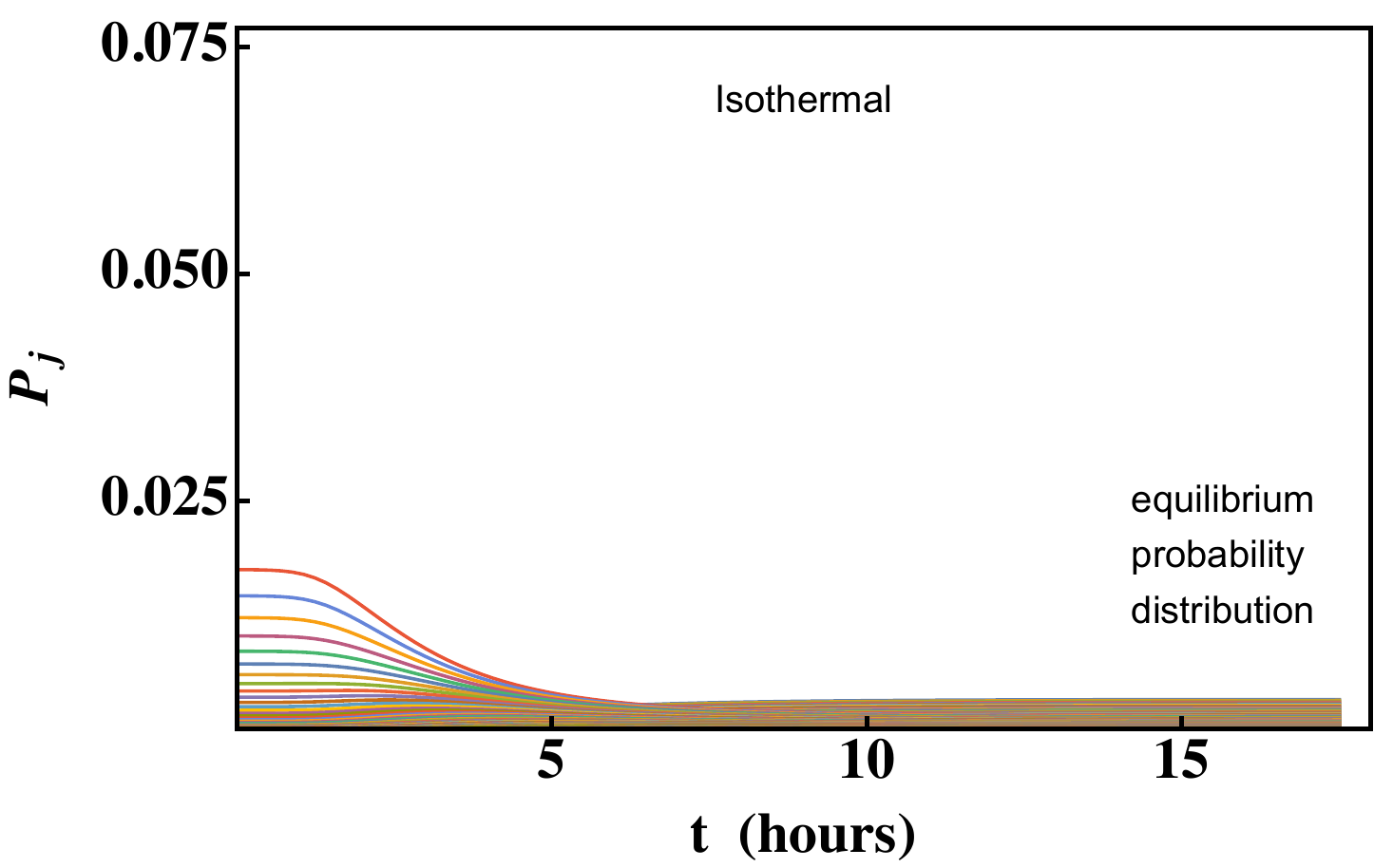}
\vspace{0.5truecm}
\includegraphics[width=0.8\textwidth]{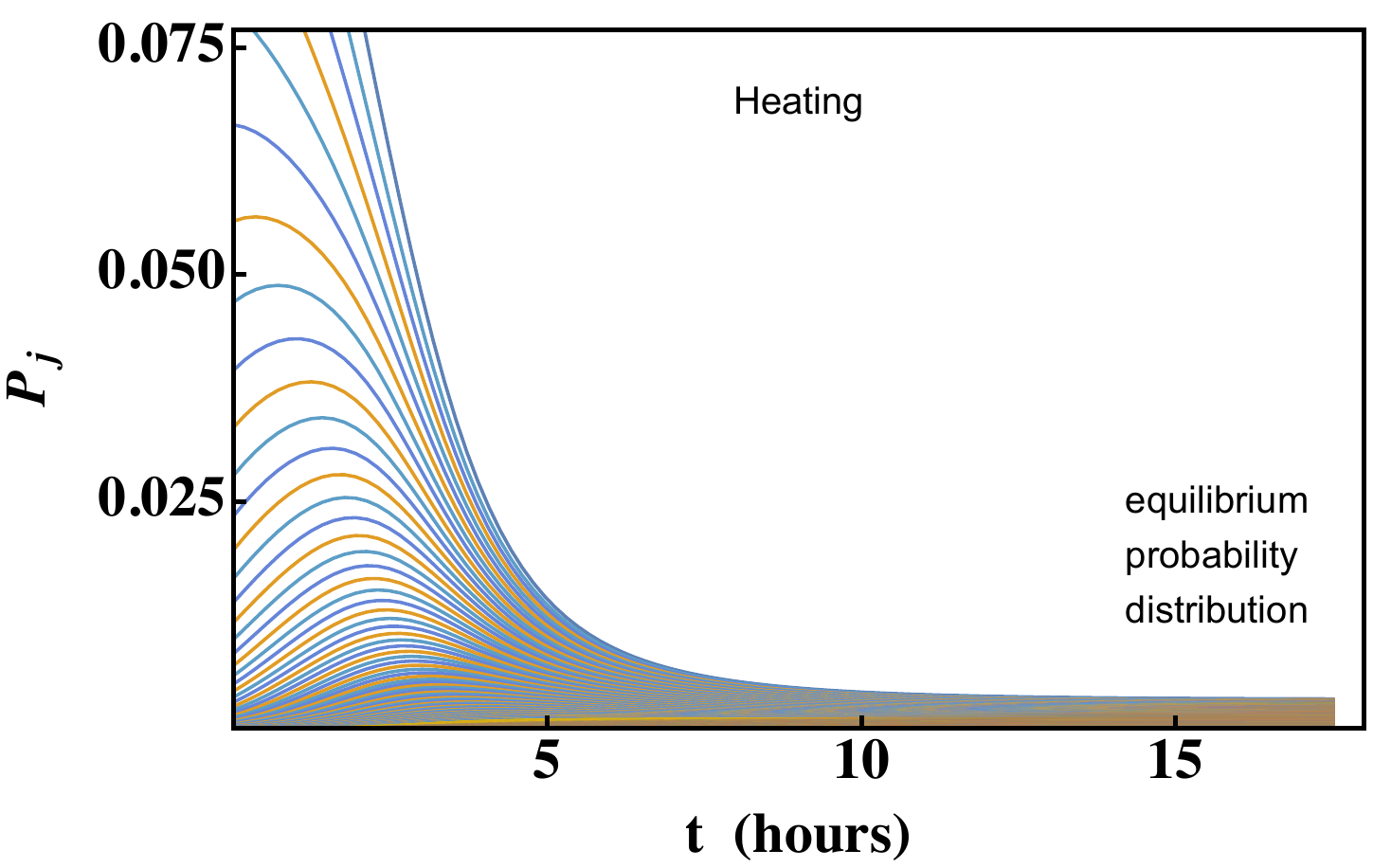}
\caption{Time evolution occupation probabilities from the initial state to stable equilibrium. The individual curves represent the occupation probability for every tenth energy eigenlevel. The image on the left is for the case of isothermal annealing at 523 K, while that on the right is for annealing with heating from 77 K to 523 K.}
\label{fig:OccupationalPvsTime}
\end{center}
\end{figure*}

The evolution in time of the occupational probabilities for all the system energy eigenlevels is shown in Figure~\ref{fig:OccupationalPvsTime} for both the isothermal annealing and annealing with heating cases. Each curve in this Figure represents the time evolution of a single each energy level from the initial state to equilibrium (only 10\% of the energy levels are plotted to reduce clutter). In both isothermal and heating processes, the occupation probability of an energy level ebbs and flows with time as defects rearrange (in the isothermal case) or appear and rearrange (in the heating case) and the system gradually approaches equilibrium.

\subsection{Maximum Entropy Production \label{results:Sproduction}}

As noted in the Introduction and in Section~\ref{REWL_sec:level2}, the intrinsic force driving the kinetics of the annealing process in the SEAQT framework are the energy transitions between energy eigenlevels guided by the SEA principle and captured by the rate of entropy production or generation, $\dot{\sigma}$. For the case of isothermal annealing, the entropy, $\langle S \rangle$, of the the PtSe$_2$ film (i.e., subsystem $A$) is calculated as an expectation value from the probability distribution and consequently evolves according to the equation of motion. The rate of entropy production for the isothermal case is calculated from the time derivative of the entropy of subsystem $A$.  For the case of annealing with heating, the entropy of subsystem $A$ includes entropy production from defect rearrangements as well as entropy brought into the subsystem via its heat interaction with the reservoir. Thus, the rate of entropy production for this case is calculated from the time derivative of the entropy of subsystem $A$ minus the entropy flowing into the film from the reservoir.

\begin{figure*}[h!]
\begin{center}
\includegraphics[width=0.49\textwidth]{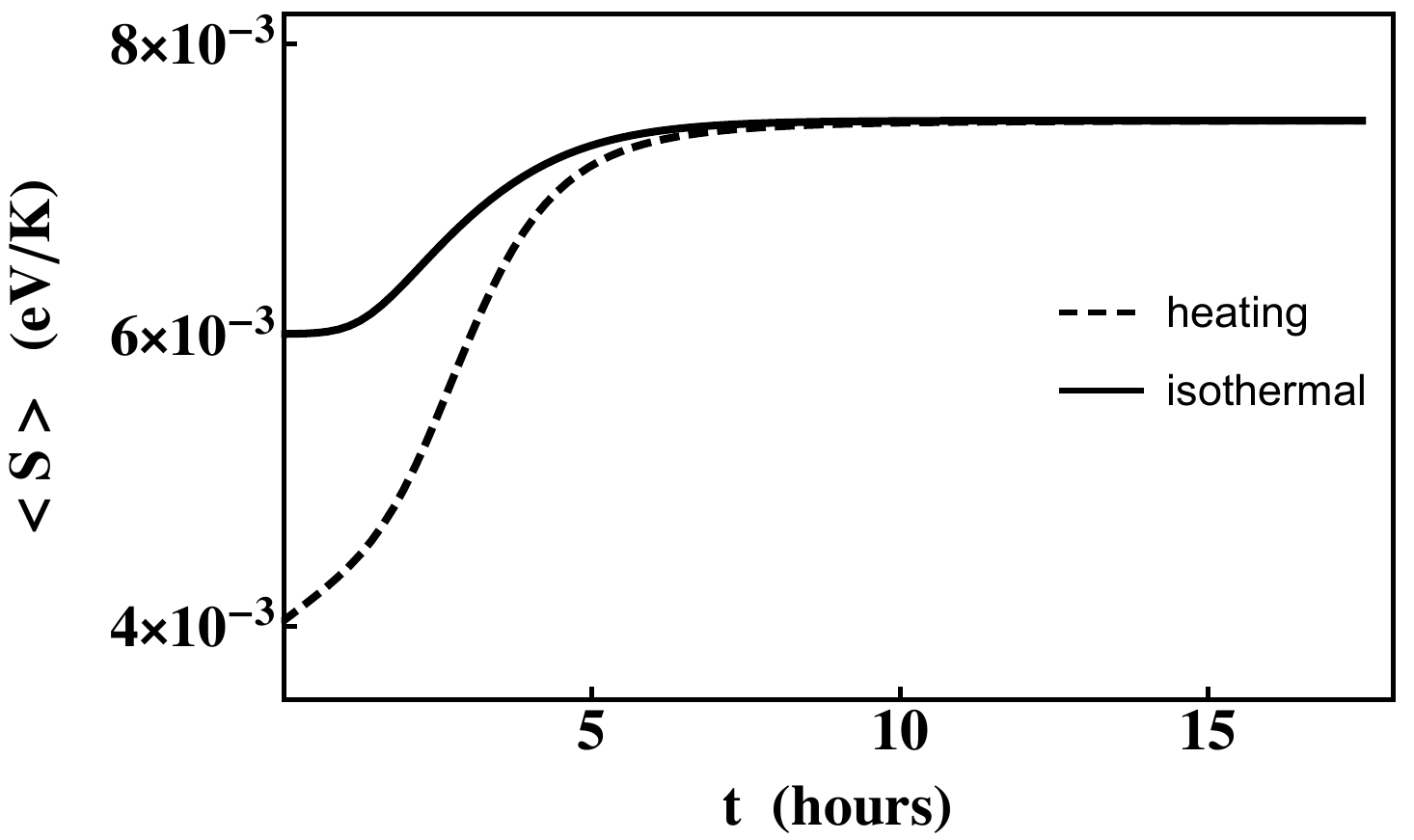}
\includegraphics[width=0.49\textwidth]{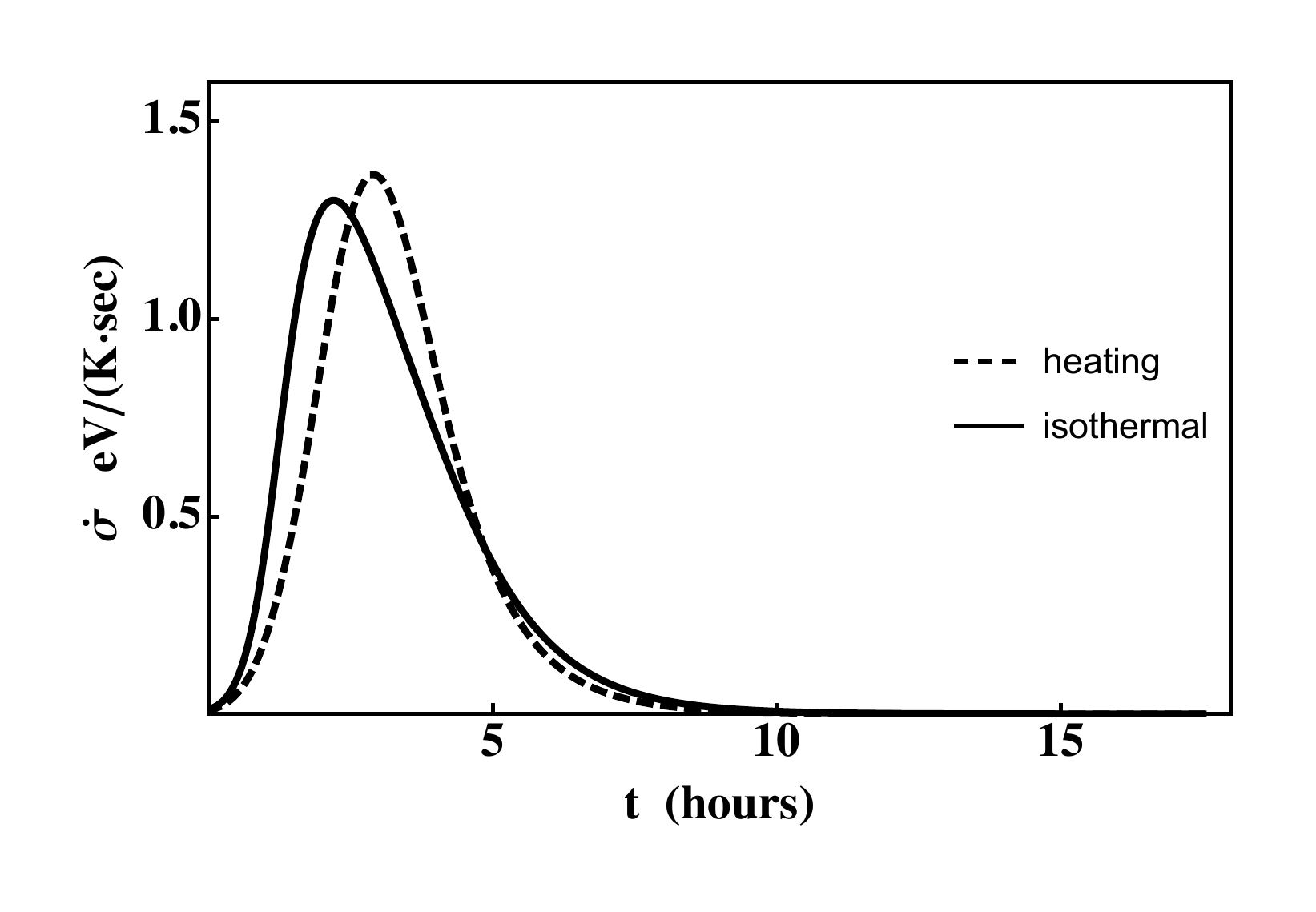}
\caption{(Left) Entropy as a function of time for the two annealing processes. The solid curve represents isothermal (constant $E$) annealing at 523 K, while the dashed curve represents annealing with heating from 77 K to 523 K. (Right) The rate of entropy production in PtSe$_2$ (subsystem A) as a function of time for two annealing processes. The solid curve represents isothermal annealing at 523 K, while the dashed curve represents annealing with heating from 77 K to 523 K.}
\label{fig:SandSprodVst}
\end{center}
\end{figure*}

Figure~\ref{fig:SandSprodVst} shows how $\langle S \rangle$ and $\dot{\sigma}$ evolve over time for the two annealing processes. Entropy increases more noticeably during heating than isothermal annealing, but much of the former increase is attributable to the heat interaction between the film and reservoir. In both annealing processes, the entropy production, $\dot{\sigma}$, which drives the kinetics, increases from zero as annealing starts, passes through a maximum as entropy changes most rapidly, then decreases to zero as entropy approaches its equilibrium (maximum) value. Maximal entropy (at constant energy) is equivalent to more commonly used conditions for equilibrium that are based upon minimal energy (at constant entropy).

\subsection{Predicted defect densities \label{results:DefectDensities}}

\begin{figure*}[h!]
\begin{center}
\includegraphics[width=0.49\textwidth]{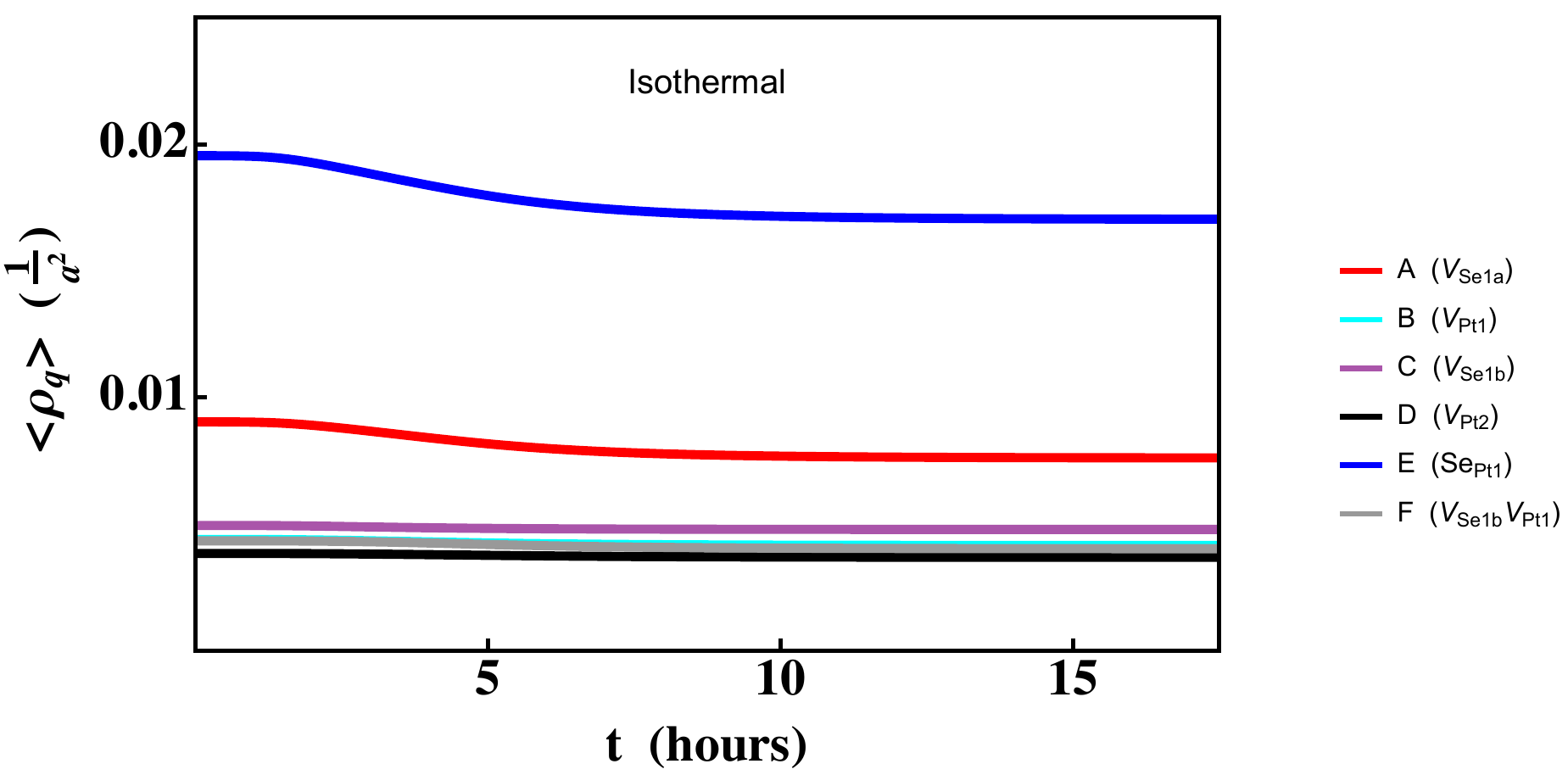}
\vspace{0.5truecm}
\includegraphics[width=0.49\textwidth]{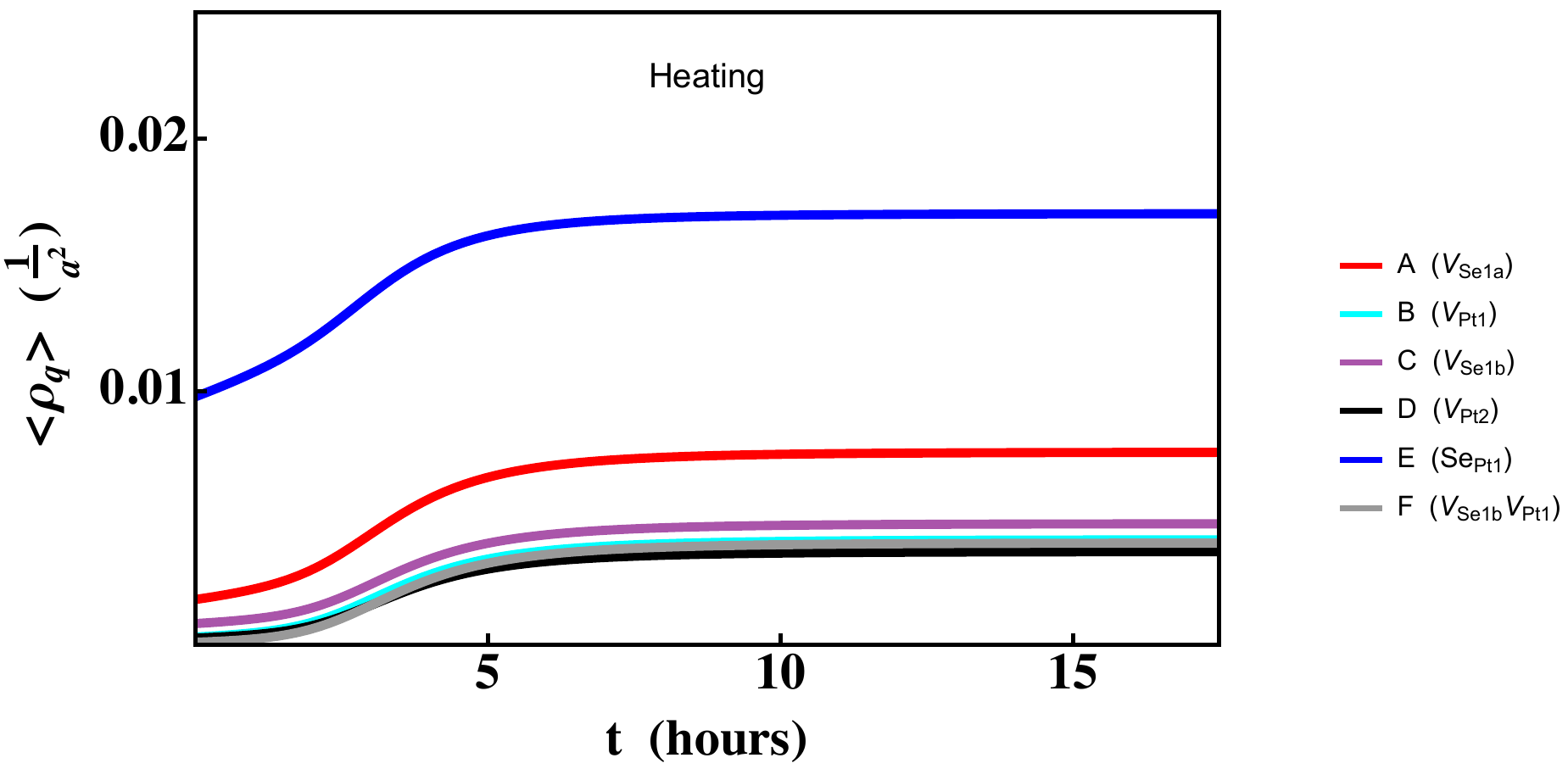}
\caption{Weighted-average defect densities versus time obtained from the time-dependent occupation probabilities predicted by the SEAQT equation of motion for (left) isothermal annealing and (right) annealing while heating.}
\label{fig:WeightedAveDefects}
\end{center}
\end{figure*}

A time-dependent picture of microstructural evolution during annealing is constructed by linking the system energy (an expected valued obtained from the time-dependent occupation probabilities) to a representative configuration with the same energy selected from among all the configurations with the same energy.  Many configurations with the same energy can have radically different appearances, so randomly selected configurations along the kinetic path in state space would produce visually disconnected microstructures. However, a visually smooth evolution of microstructure can be ensured by selecting a representative configuration at each time with microstructure descriptors that are tied to the initial configuration~\cite{2022MvSR}.  In the present case, the microstructures along the kinetic path are selected at each instant of time from representative microstructures that have defect arithmetic average densities (see Figure~\ref{fig:ArithmeticAve}) of the $E$, $A$, $C$, $B$, $F$, and $D$ defects (in that order) with eigenenergies similar to the eigenenergies used to calculate the expected value of the energy. These eigenenergies correspond to the non-zero occupation probabilities predicted at each instant of time, probabilities which are then used to determine the expected (weighted-average) value for each defect density. This procedure guarantees that these weighted-average values vary smoothly with time during both the isothermal annealing and annealing with heating processes as seen in Figure~\ref{fig:WeightedAveDefects}. The left side of each graph in Figure~\ref{fig:WeightedAveDefects} corresponds to the initial configuration of defects at the start of the annealing processes, and the right side of the graphs represent the stable equilibrium configurations achieved at the end of the anneals.

As noted previously, the energy remains constant during isothermal annealing so that the weighted-average densities of the defects also remain relatively unchanged (left panel of Figure~\ref{fig:WeightedAveDefects}). In the heating case, energy increases as energy in a heat interaction is transferred from the thermal reservoir to the 2D film so that the weighted-average densities of the defects increase noticeably (right panel of Figure~\ref{fig:WeightedAveDefects}). The density of all the defect types nearly double from the initial state to stable equilibrium. Nevertheless, the density of each defect type at the final stable equilibrium state for both the isothermal and heating cases are the same as they should be at a common final temperature. 

A final note is that the equilibrium densities predicted here are comparable to those seen experimentally in Table 1 of \cite{Zheng2019}. For both the isothermal and heating cases, the values in Figure 9 at equilibrium for a lattice constant of 0.375 nm [3] agree with the experimental densities in \cite{Zheng2019} to the same order of magnitude. In fact, for D and E, our predicted values are within the error bars for the experimental values, while those for A, B and C differ by a factor of about 2. The reason that the density from one defect type to the other differs is primarily due to the defect formation energies being different although from the isothermal case shown in Figure 9, it is noted that the defect densities are also affected by the interactions among groups of defects. 

\subsection{Predicted defect configurations \label{results:Configurations}}

\begin{figure*}[h!]
\begin{center}
\includegraphics[width=0.49\textwidth]{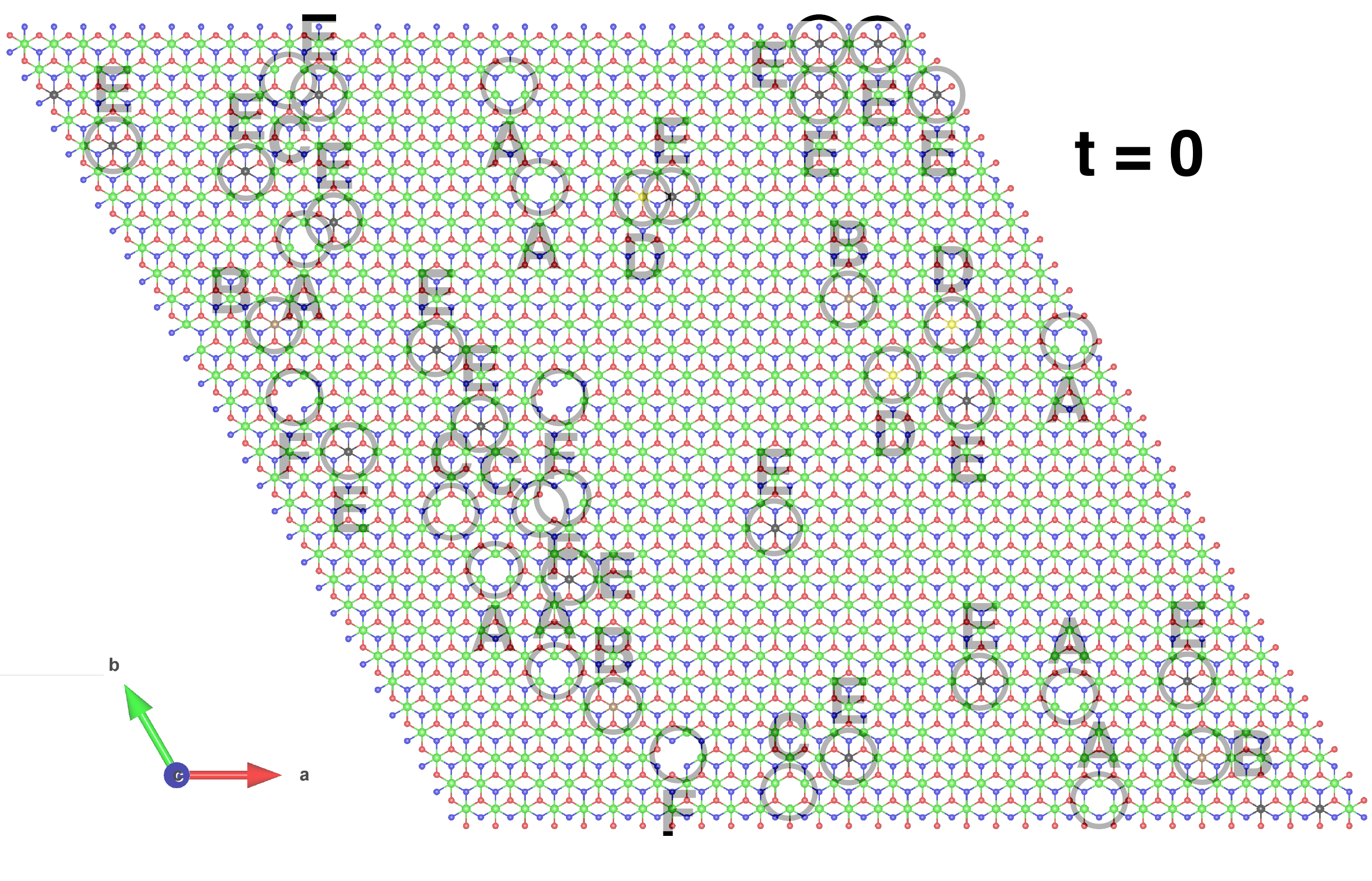}
\includegraphics[width=0.49\textwidth]{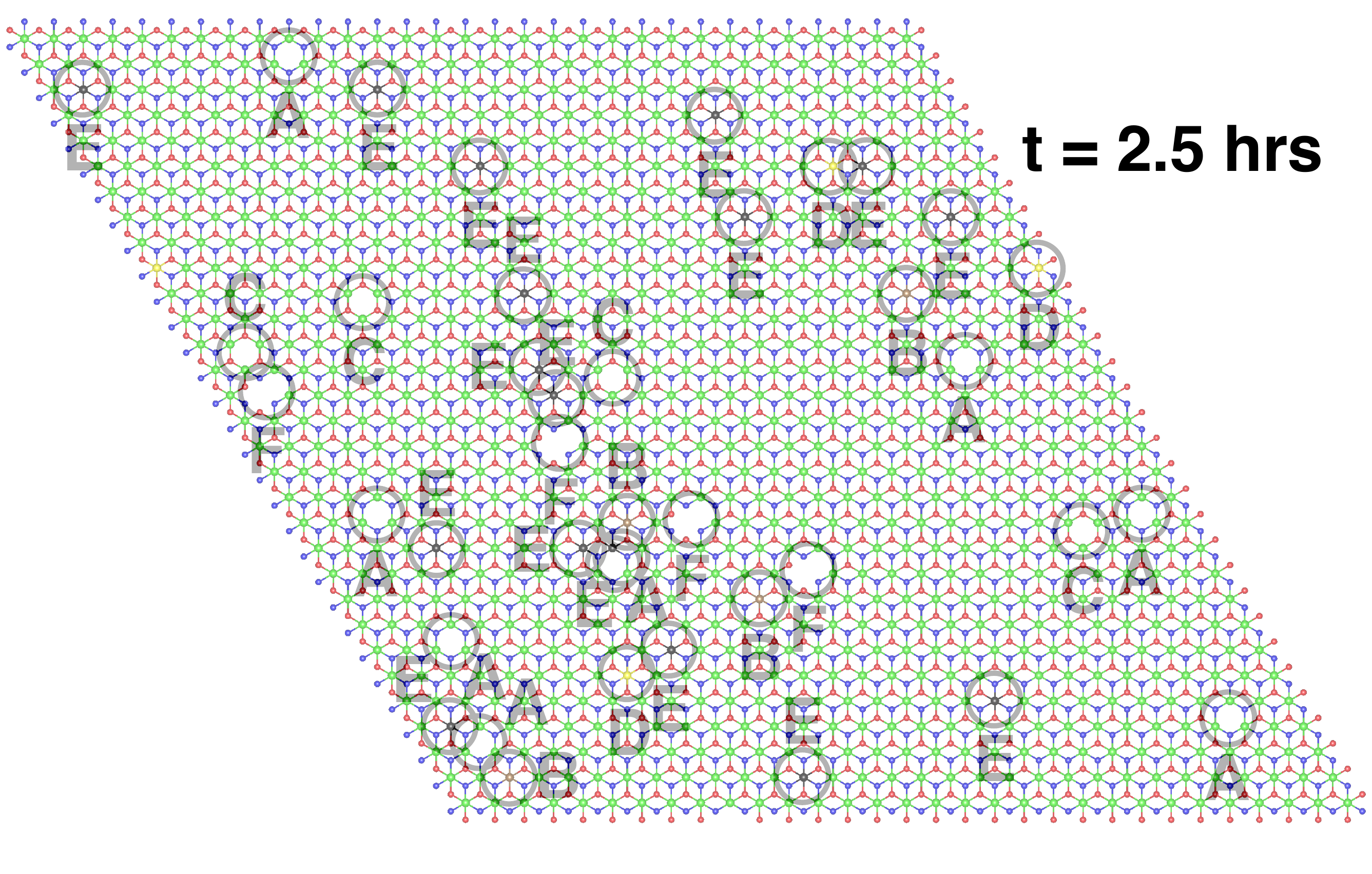}
\includegraphics[width=0.49\textwidth]{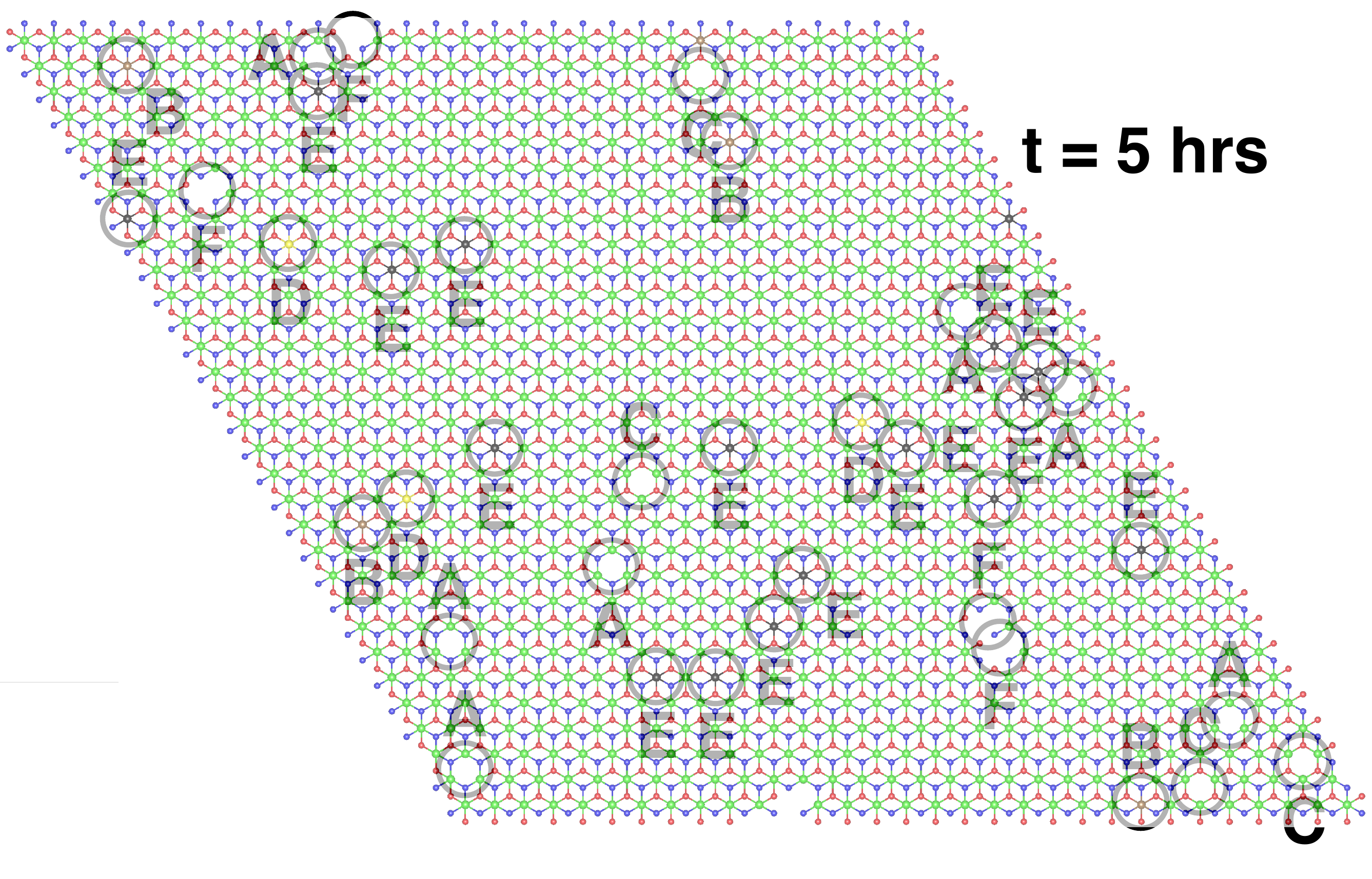}
\includegraphics[width=0.49\textwidth]{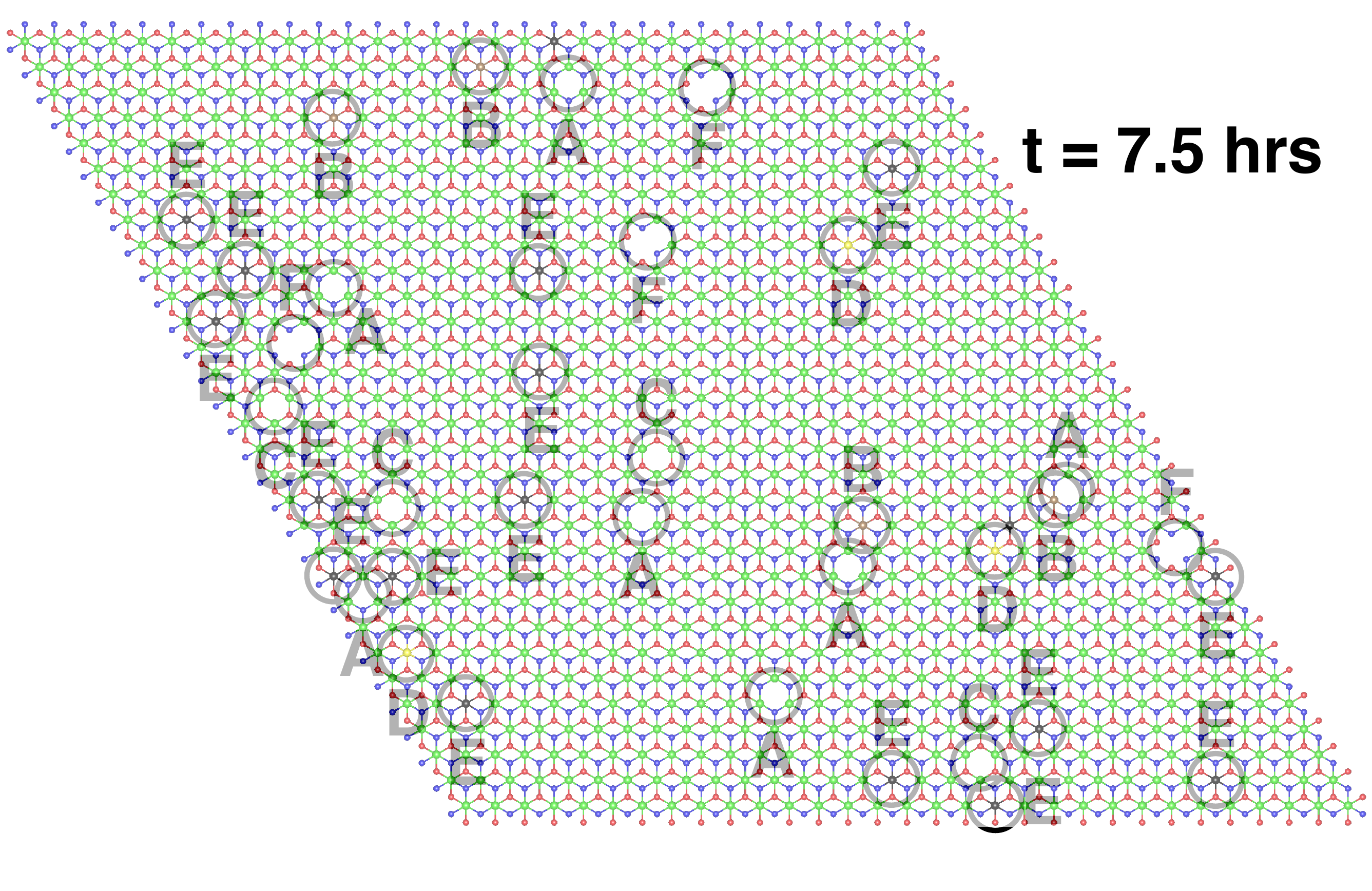}
\includegraphics[width=0.49\textwidth]{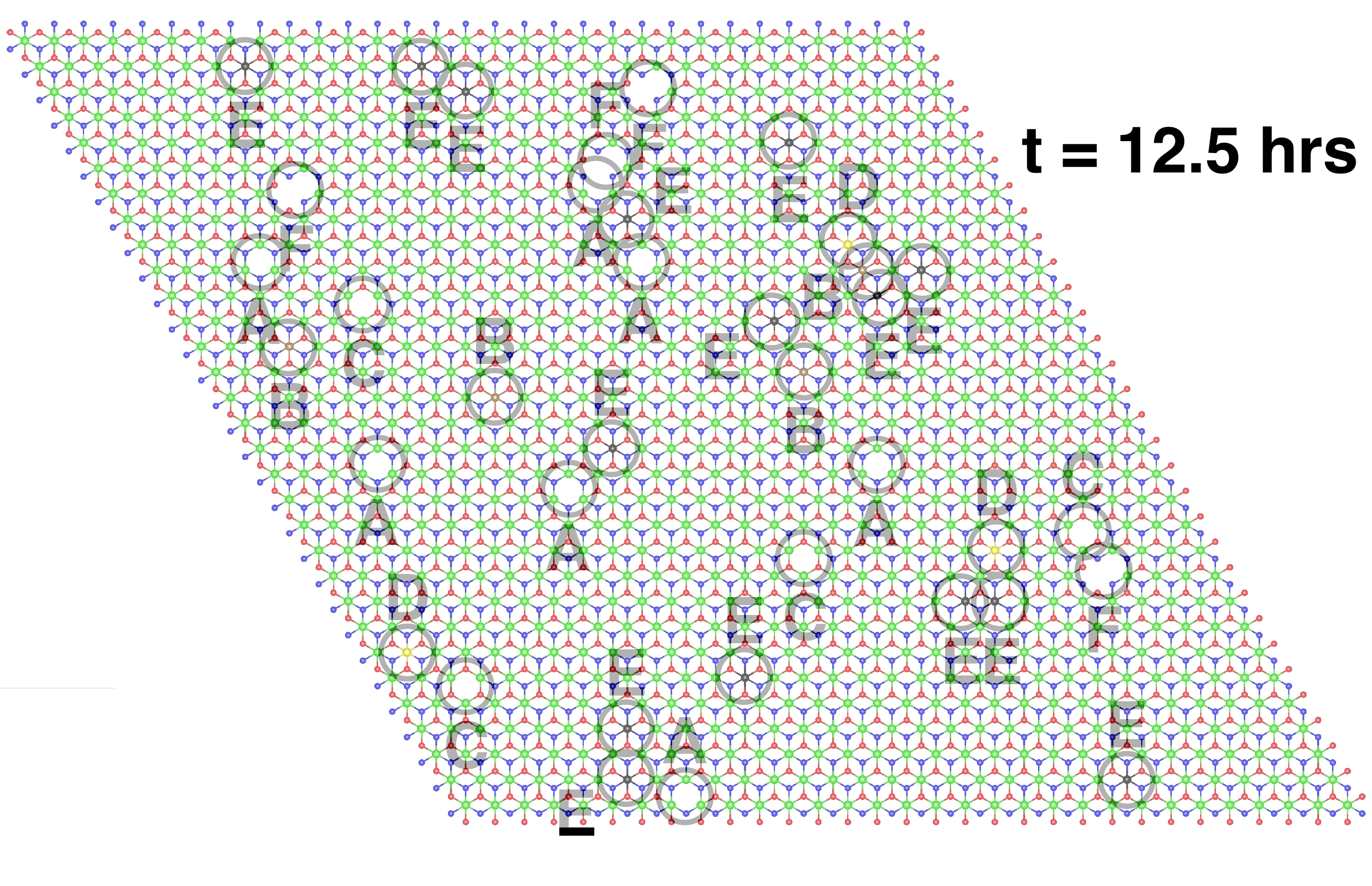}
\includegraphics[width=0.49\textwidth]{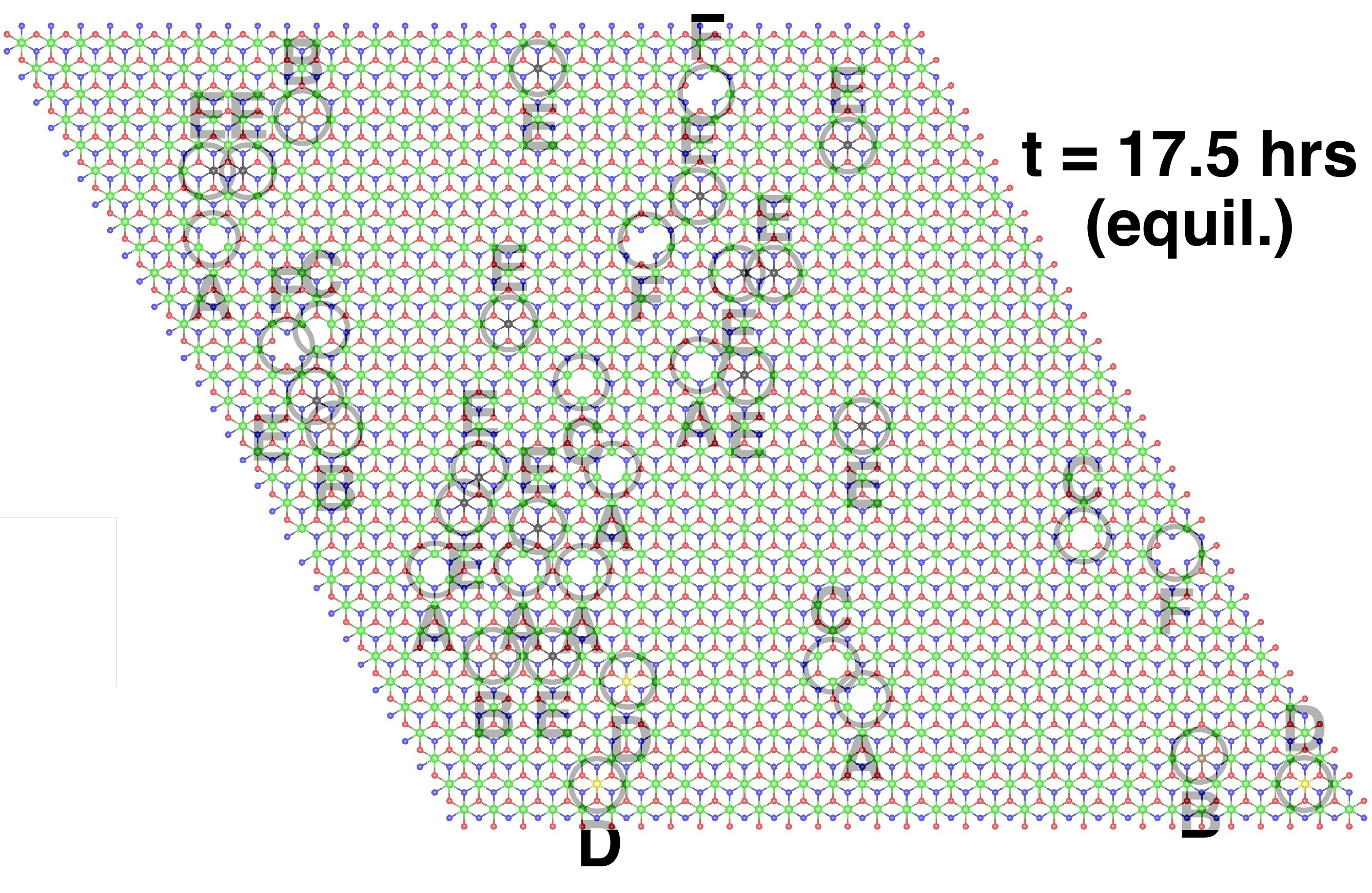}
\caption{Evolution of defect configurations from the initial state to stable equilibrium during isothermal annealing at 523 K. Defects less than $5\,a$ apart interact energetically.}
\label{fig:TimeEvolutionConfigsIsothermal}
\end{center}
\end{figure*}

\begin{figure*}[h!]
\begin{center}
\includegraphics[width=0.49\textwidth]{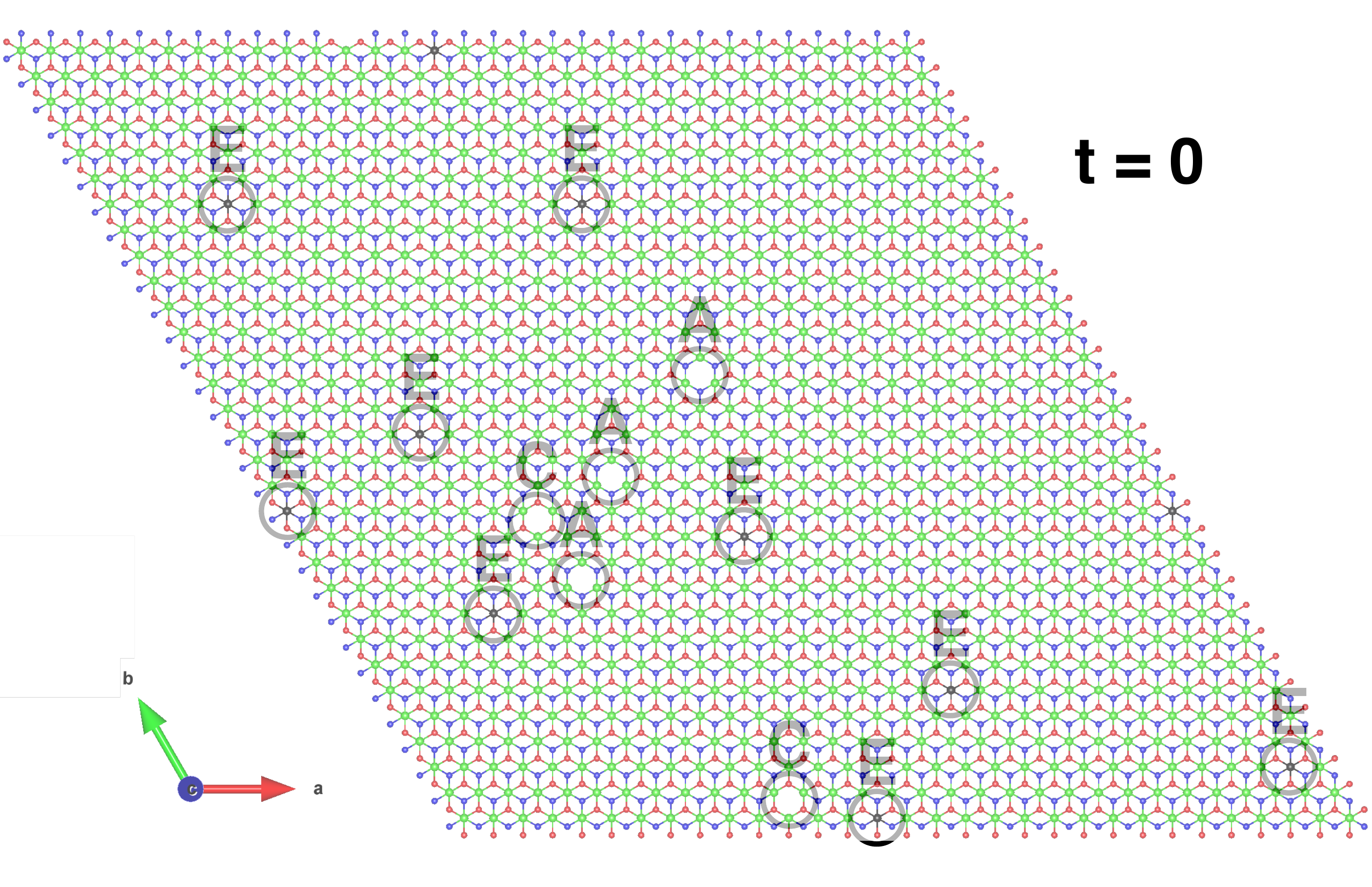}
\includegraphics[width=0.49\textwidth]{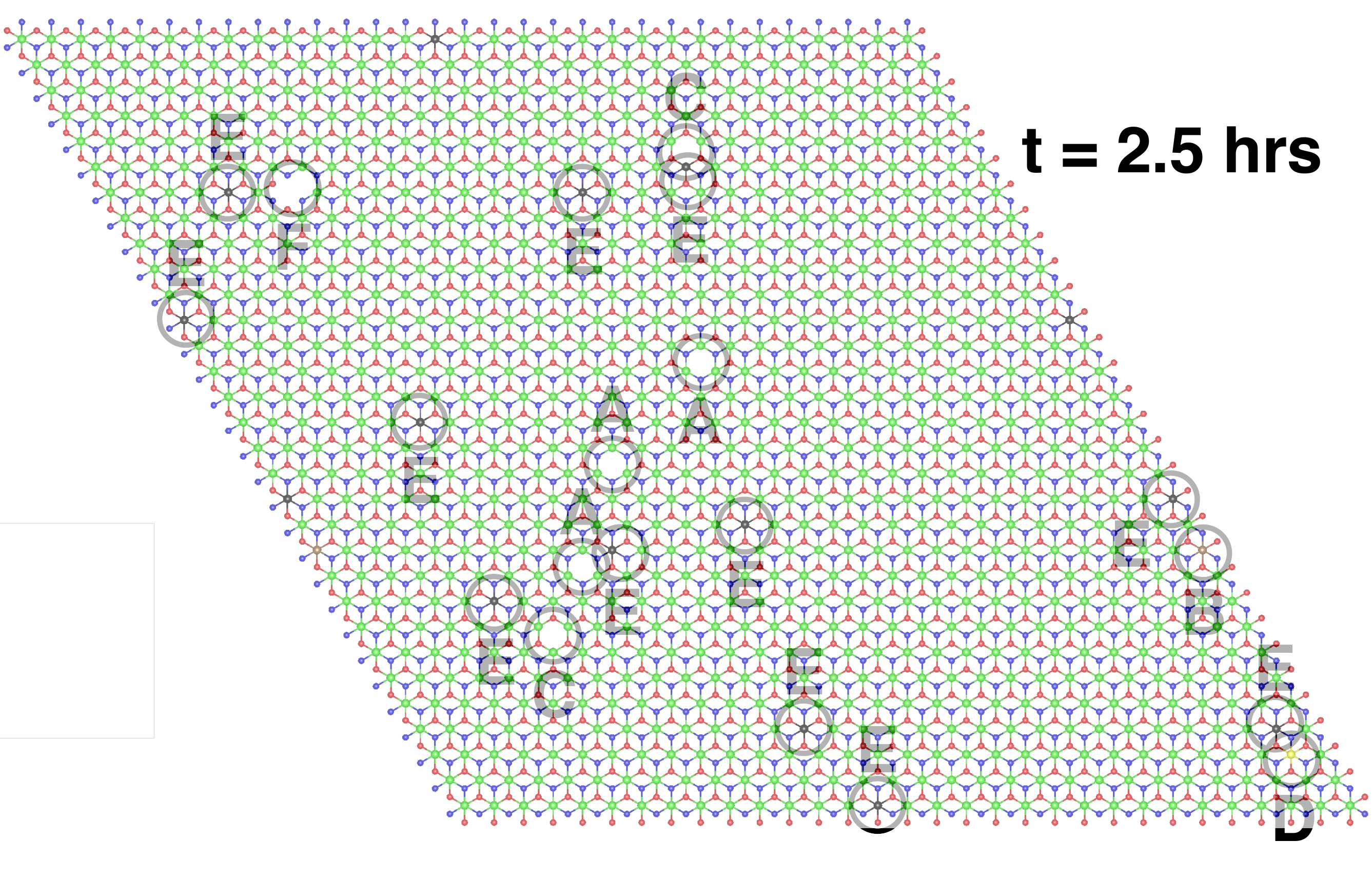}
\includegraphics[width=0.49\textwidth]{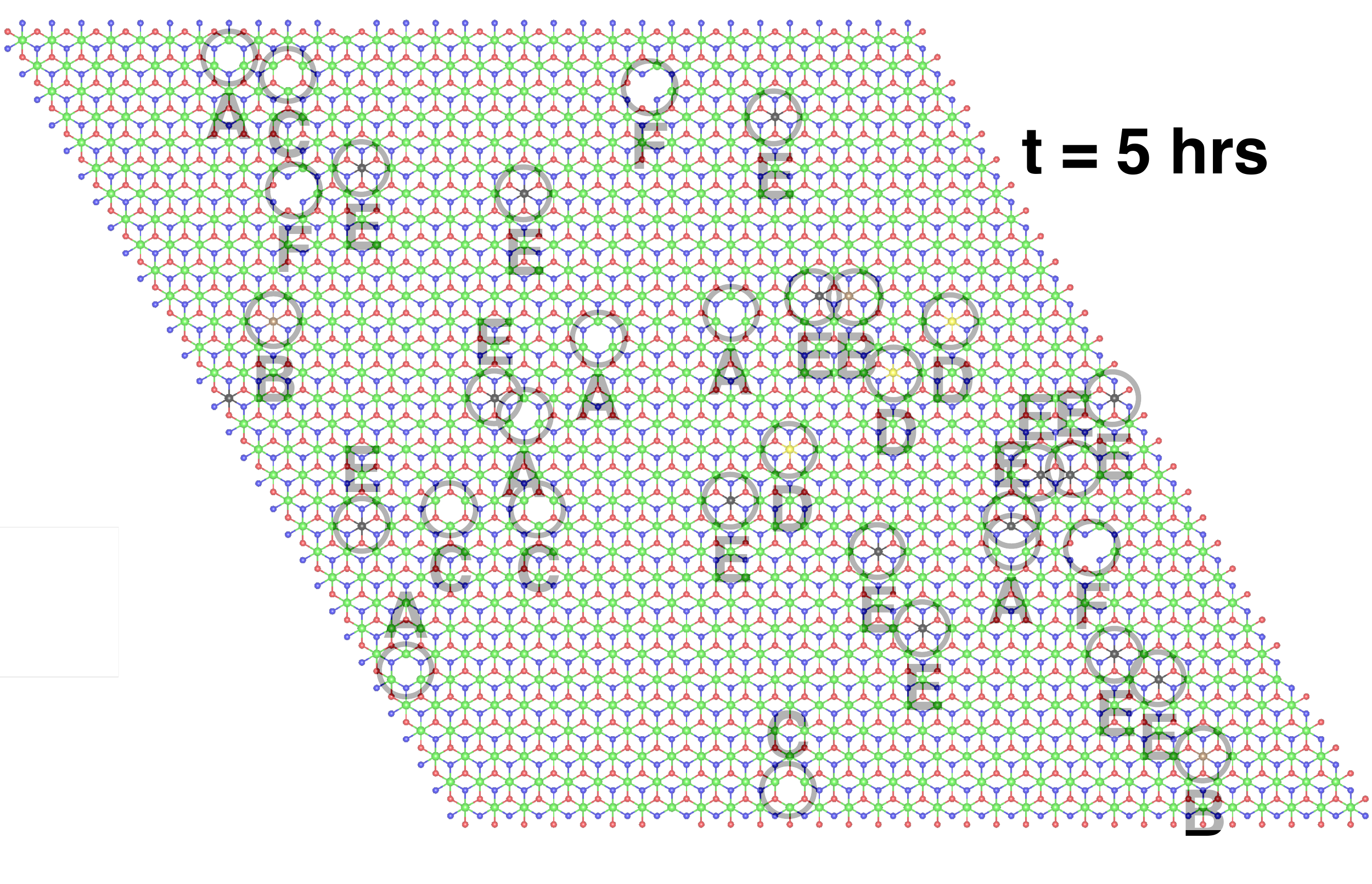}
\includegraphics[width=0.49\textwidth]{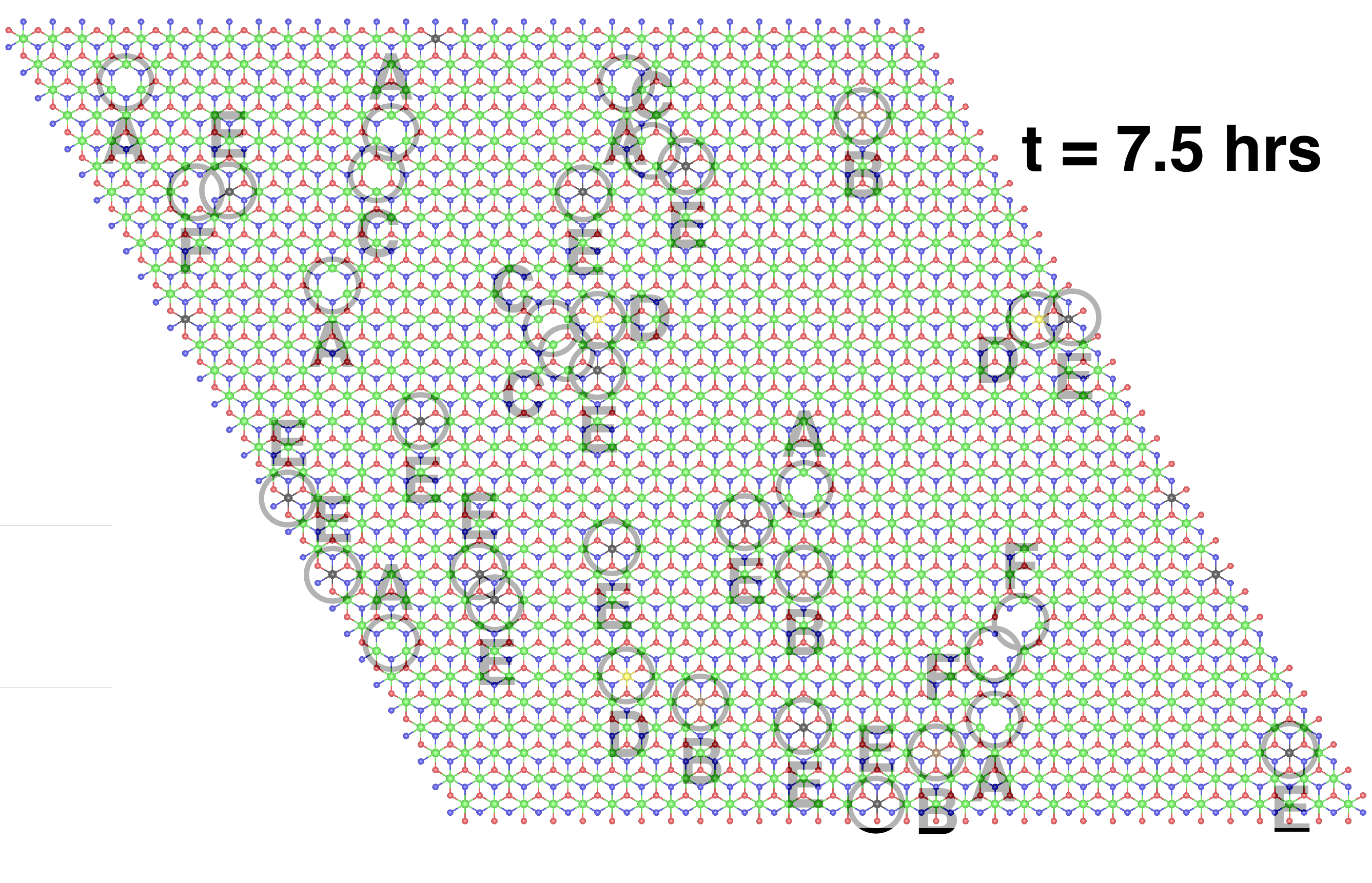}
\includegraphics[width=0.49\textwidth]{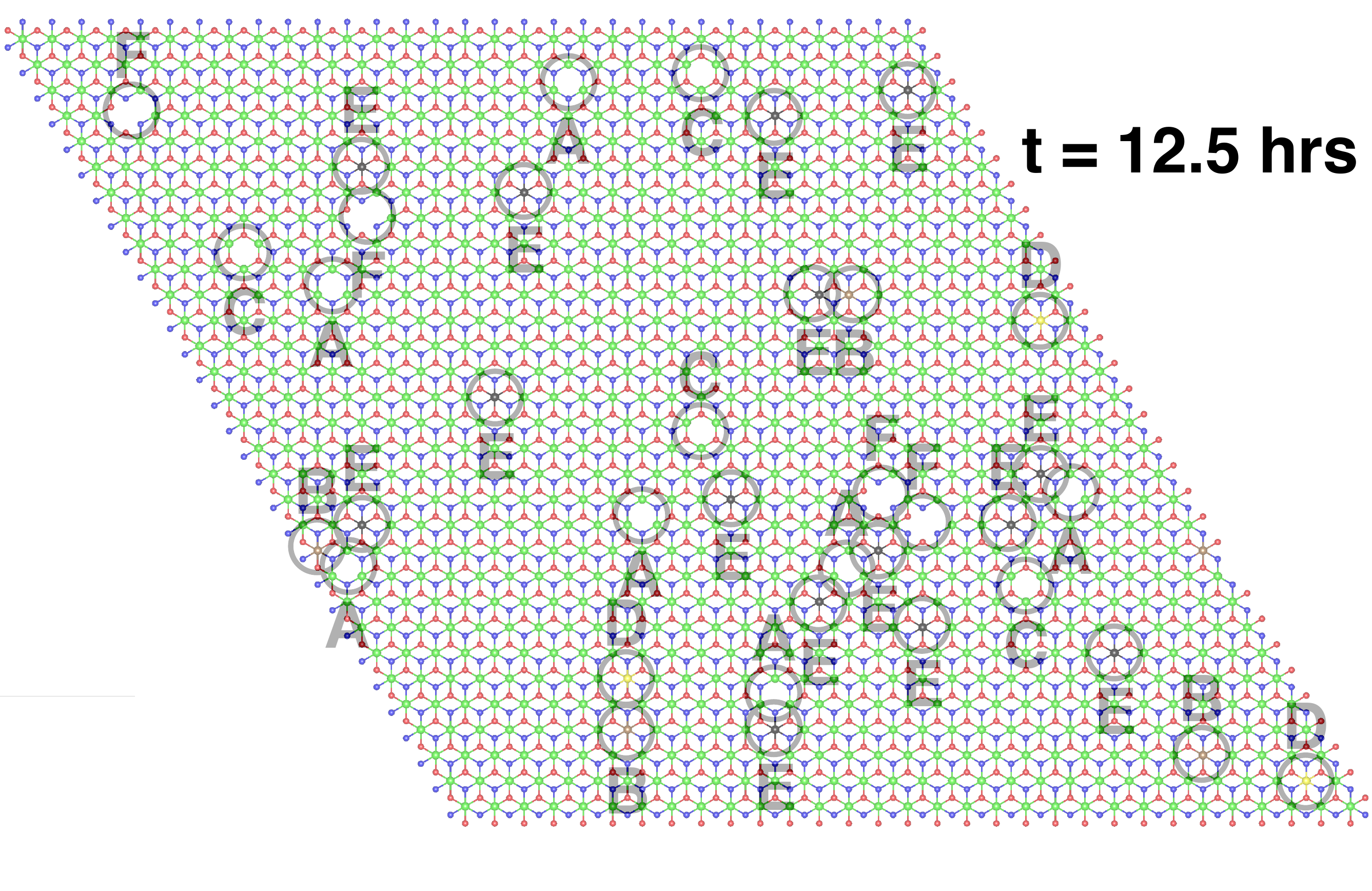}
\includegraphics[width=0.49\textwidth]{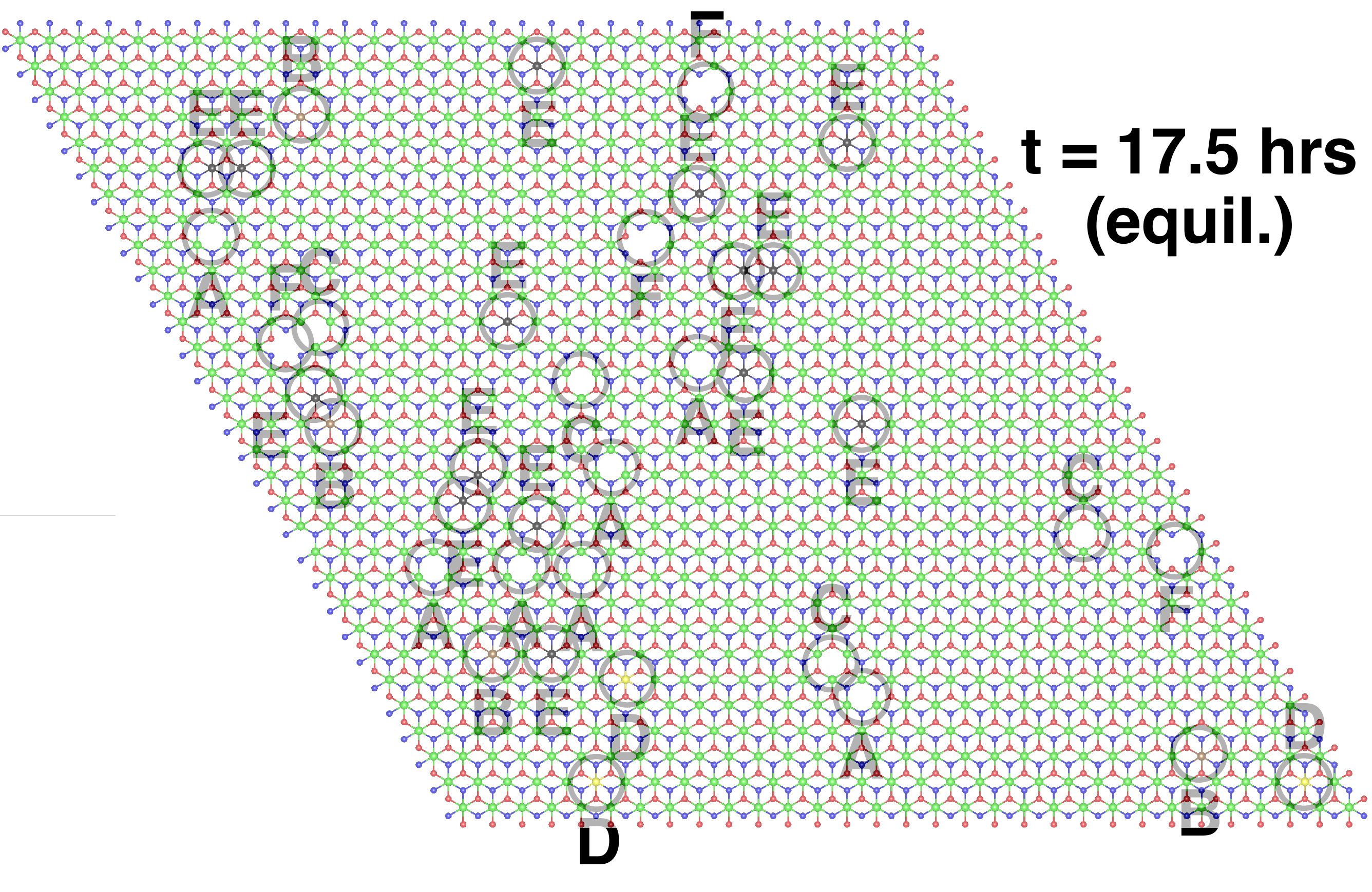}
\caption{Time evolution of defect configurations from the initial state to stable equilibrium during annealing while heating from 77 K to 523 K.  The lower defect density at $t=0$ reflects the lower energy of the system when annealing begins at 77 K. Defects less than $5\,a$ apart interact energetically; most defects are isolated at $t=0$, and almost all of them interact with others for annealing times beyond $t=2.5$ hr. }
\label{fig:TimeEvolutionConfigsHeating}
\end{center}
\end{figure*}

Figure~\ref{fig:TimeEvolutionConfigsIsothermal} shows SEAQT-predicted microstructures at six times during the isothermal (constant energy) annealing process at 523 K. The spatial dimension along each side of the model lattice represents $30\,a$ or $11.3$ nm. For comparison, the STM images of Figure~\ref{DefectMigration} are about $9.5$ nm on a side. The time $t=0$ represents the initial configuration with defect densities based on STM images. The annealing times of the second and fourth images (2.5 and 7.5 hr, respectively) in Figure~\ref{fig:TimeEvolutionConfigsIsothermal} correspond to the STM images of Figure~\ref{fig:TimeEvolutionConfigsIsothermal}$a$ and $b$ (2.5 hr) and $c$ and $d$ (7.5 hr). Consistent with the defect densities seen in the experimental data, the initial configuration chosen for the SEAQT evolution for isothermal annealing is, as seen at $t=0$ in Figure~\ref{fig:TimeEvolutionConfigsIsothermal}, a mix of primarily interacting defects with a few isolated defects included. During isothermal (i.e., constant energy) annealing, the SEAQT equation of motion predicts the rearrangement of these defects with a resultant increase in system entropy as well as  energy. The latter is offset by an equivalent decrease in energy due to the elimination of three E defects and one A defect so that the system energy remains constant. The number of other defects remains the same. This is consistent with the evolution of the weighted-average defect densities seen in the left panel of Figure~\ref{fig:WeightedAveDefects}. Of course, the increase in entropy is not only due to the defect rearrangements but also to an increase in the number of interacting defects with a consequent decrease in the number of isolated defects. This is clearly seen in Figure~\ref{fig:TimeEvolutionConfigsIsothermal} at 2.5 hr and continues until equilibrium is reached at 17.5 hr. During this process, defects agglomerate more and more and move toward one another and the interacting distance between defects decreases to less than $5\,a$, which is again consistent with experimental results. Also, the defect configuration at equilibrium predicted by the SEAQT framework after 17.5 hours of annealing is compatible with the STM images taken after 7 annealing periods with the equivalent accumulated time. Note that the SEAQT predicted configurations for isolated defects do look somewhat different than the STM images, but, as explained above, the STM images are based on the local electron density of the surface atoms. Thus, for example, in the case of a missing Pt atom in the top Pt layer of a film two $\mbox{PtSe}_{2}$ layers thick (defect B), {\em three} bright dots are seen in the STM images. In contrast, B defects in Figure~\ref{fig:TimeEvolutionConfigsIsothermal} (and Figure~\ref{fig:TimeEvolutionConfigsHeating}) are represented schematically by individual labeled circles positioned where each defect is located in the film.

Representative defect configurations during heating from $77$ K to $523$ K are shown in Figure~\ref{fig:TimeEvolutionConfigsHeating}.  For this annealing process while heating the film, the low starting temperature implies an initially low number of defects (14 total at the initial annealing time of $t=0$). The initial state at $t=0$ is chosen based on an initial system energy compatible with 77 K. Using an STM image developed in \cite{Zheng2019}, an initial configuration consistent with this energy is constructed using only A, C, and E defects. About half of these defects interact energetically with other defects (because they lie less than $5\,a$ away from other defects), and half are isolated defects. The number of defects increases as energy in a heat interaction flows from the reservoir into the film and the defects arrange into clusters of interacting defects. After 2.5 hr, there are 20 defects and all of them are interacting with neighboring defects --- there are no isolated defects. The 36 defects present after 7.5 hr is nearly the stable equilibrium value (see right side of Figure~\ref{fig:WeightedAveDefects}). As for the isothermal case, from 7.5 hr onward there is only one isolated defect within the configuration. As expected, the equilibrium configurations predicted for both the isothermal and heating cases are identical even though the evolutions are completely independent of each other.

Note that because the arrangements of interacting defects in Figures \ref{fig:TimeEvolutionConfigsIsothermal} and \ref{fig:TimeEvolutionConfigsHeating} are complex, no obvious trend in how defects in different layers affect defect ordering was seen. This is probably due to the fact that all of the defect arrangements considered lowered the film energy by similar amounts relative to the same number of widely separated defects. Furthermore, because the interacting defect groups had different interaction energies, no obvious correlations among specific types of defects were observed.

Finally, the comparisons shown here of the experimental defect configurations of Figure~\ref{DefectMigration} with the configurations predicted by the SEAQT equation of motion during annealing (either isothermal or heating) clearly demonstrate that the SEA principle correctly captures the appearance of defect patterns as well as the kinetics that form them. It does so based solely on the kinetics of the energy landscape energy transitions guided by the SEA principle.

\section{Conclusions \label{conclusions_sec:level1}}
As demonstrated above, the SEAQT framework is able to predict defect configurations in 2D PtSe$_2$ following annealing at 523 K that are consistent with experimentally observed configurations after 17.5 hours. The predictions include both the observed number density of all the observed defects as well as a qualitative assessment of the spatial arrangement of these defects. Furthermore, fitting the SEAQT annealing kinetics from an initial configuration to the final stable equilibrium configuration to the STM observations could in this case be done using a constant SEAQT relaxation parameter, $\tau$. 

Also, demonstrated here is that the number densities of the defect types are effective microstructural descriptors for linking the kinetic evolution in state space to a physically realistic picture of the defect configurations in 2D PtSe$_2$. Linking the state space results inherent to the SEAQT framework to microstructural changes is an important addition to this framework since it provides a bridge between the thermodynamic property predictions and the structural changes of the system. 

Another important conclusion is that of the advantage afforded by only having to use a single general kinetic principle to predict the phenomena present in any process. The SEAQT framework does this by using the SEA principle to guide the predictions of the kinetics of energy landscape energy transitions. It is these kinetics that form the underlying basis for the phenomena present in any process. Such a single principle avoids the need to guess, as often occurs, at what the correct phenomenological kinetic mechanisms are for single and particularly coupled phenomena. It furthermore avoids the problem of having to have separate models for transitions from one type of phenomenon to another that require a different phenomenological mechanism for each. 

Finally, a significant advantage of the SEAQT framework is its reduced computational overhead when compared to conventional approaches. Creating the energy landscape requires the largest computational resources. In the present application, this involves the use of a non-Markovian Monte Carlo approach. However, this approach is only used to establish the landscape and is not used to predict the kinetics of the process and, thus, only has to be done once. To establish the kinetics of the process, the SEAQT equation of motion is used but this only involves simultaneously solving a large number of first-order ordinary differential equations in time. Doing so requires the use of very moderate computational resources.

\begin{acknowledgments}
The authors acknowledge Chenggang Tao for sharing his scanning tunneling microscopy results and Advanced Research Computing at Virginia Tech for providing computational resources and technical support that have contributed to the results reported within this paper.
\end{acknowledgments}

\begin{table*}[h!]
\begin{tabularx}{0.85\textwidth} { 
  | >{\raggedleft\arraybackslash}c 
  | >{\raggedright\arraybackslash}c 
  | >{\raggedright\arraybackslash}X 
  | >{\raggedright\arraybackslash}X | }
  \hline
\multicolumn{1}{|c|}{ Defect label } & \multicolumn{1}{|c|}{Defect type} & \multicolumn{1}{|c|}{Defect description} & \multicolumn{1}{|c|}{Effect on concentration} \\ \hline
$V_{\mbox{\scriptsize Se1a}}$ & $A$ (Potts spin $=1$) & Top Se vacancy in 1st layer & Removes one Se atom \rule[-.3\baselineskip]{0mm}{5mm} \\
$V_{\mbox{\scriptsize Pt1}}$ & $B$ (Potts spin $=2$) & Pt vacancy in 1st layer & Removes one Pt atom \\
$V_{\mbox{\scriptsize Se1b}}$ & $C$ (Potts spin $=3$) & Bottom Se vacancy in 1st layer & Removes one Se atom \\
$V_{\mbox{\scriptsize Pt2}}$ & $D$ (Potts spin $=4$) & Pt vacancy in 2nd layer & Removes one Pt atom \\
$\mbox{Se}_{\mbox{\scriptsize Pt1}}$ & $E$ (Potts spin $=5$) & Se on 1st layer Pt site (anti-site) & Removes one Pt atom, adds one Se atom \\
$V_{\mbox{\scriptsize Se1b}}V_{\mbox{\scriptsize Pt1}}$ & $F$ (Potts spin $=6$) & Pt vacancy in 1st layer and bottom Se vacancy in 1st layer & Removes one Pt atom, removes one Se atom \\ \hline

\end{tabularx}
\caption{Defect types and designations in 2D PtSe$_2$.}
\label{defectTypes} 
\end{table*}

\begin{appendices}

\section{Appendix \label{appendix}}

\subsubsection{Chemical Potentials \label{ChemPotentialsSubSection}}

Expressions for the chemical potentials required in Eqs.~(\ref{FormationEnergy}) and (\ref{GroupEnergy}) are obtained as follows. Both $\mu_{\mbox{\scriptsize Pt}}$ and $\mu_{\mbox{\scriptsize Se}}$ are composition-dependent, and since defects alter the composition of the film, these chemical potentials change with the numbers of defects. Considering first the perfect stochiometric compound, the total energy for a formula unit of $\mbox{Pt}\mbox{Se}_2$, which is equal to its chemical potential, is constructed from the chemical potentials of its component species via
\begin{equation} \label{chemPotentialExpression}
\mu_{\mbox{\scriptsize PtSe}_2} = E_{\mbox{\scriptsize PtSe}_2} = 2\, \mu_{\mbox{\scriptsize Se}} + \mu_{\mbox{\scriptsize Pt}}
\end{equation}
From DFT calculations, the energy of formation of stochiometric $\mbox{Pt}\mbox{Se}_2$ is
$$E_{\mbox{\scriptsize PtSe}_2} = -14.650 \; \frac{\mbox{\scriptsize eV}}{\mbox{\scriptsize formula unit}}$$. 

Film growth takes place under Se-rich conditions, and the chemical potential of Se is fixed at the lattice energy of Se, i.e., $\mu_{\mbox{\scriptsize Se}} = E_{\mbox{\scriptsize Se}} = -3.475 \; \frac{\mbox{\scriptsize eV}}{\mbox{\scriptsize atom}}$. From these calculated values for $\mu_{\mbox{\scriptsize Se}}$ and $E_{\mbox{\scriptsize PtSe}_2}$, Equation~(\ref{chemPotentialExpression}) can be rearranged to yield the chemical potential of Pt in a Se-rich film. Thus,
\begin{equation} \label{muPt}
\mu_{\mbox{\scriptsize Pt}}^{\mbox{\tiny Se-rich}} = E_{\mbox{\scriptsize PtSe}_2} - 2\, E_{\mbox{\scriptsize Se}} = \; -7.700 \; \frac{\mbox{\scriptsize eV}}{\mbox{\scriptsize atom}}
\end{equation}

Similarly, for a Pt-rich film, the chemical potential of pure Pt can be taken to be the calculated lattice energy of Pt, i.e., $\mu_{\mbox{\scriptsize Pt}} = E_{\mbox{\scriptsize Pt}} = -6.036 \; \frac{\mbox{\scriptsize eV}}{\mbox{\scriptsize atom}}$. Using this value and $E_{\mbox{\scriptsize PtSe}_2}$, Equation~(\ref{chemPotentialExpression}) can be rearranged to provide an expression for the chemical potential of Se (in a Pt-rich film). Thus,
\begin{equation} \label{muSe}
\mu_{\mbox{\scriptsize Se}}^{\mbox{\tiny Pt-rich}} = \frac{1}{2} \left( E_{\mbox{\scriptsize PtSe}_2} - E_{\mbox{\scriptsize Pt}} \right) = \; -4.307 \; \frac{\mbox{\scriptsize eV}}{\mbox{\scriptsize atom}}
\end{equation}
These values for the chemical potentials of Pt and Se in films with the extremes of the composition range (Se-rich and Pt-rich) are listed in Table \ref{RichChemPotentials}.

\begin{table*}[h!]
\centering
\begin{center}
\begin{tabular}{| c | c | c |} \hline
& Se-rich & Pt-rich \\ \hline
$\mu_{\scriptsize \mbox{Se}}$  & $-3.475$ & 
$-4.307$ \rule[-.3\baselineskip]{0mm}{6mm} \\[3mm]
\hline
$\mu_{\scriptsize \mbox{Pt}}$  & $-7.700$ & 
$-6.036$ \rule[-.3\baselineskip]{0mm}{6mm} \\[3mm] \hline
\end{tabular}
\end{center}
\caption{Limiting chemical potentials in 2D PtSe$_2$ (expressed in $\mbox{eV}/\mbox{atom}$).}
\label{RichChemPotentials} 
\end{table*}

The chemical potentials needed to calculate the defect formation or group energies (Eqs. (\ref{FormationEnergy}) and (\ref{GroupEnergy}), respectively for any film composition between these concentration limits is calculated from linear interpolation via
\begin{align}
\mu_{\mbox{\scriptsize Se}} = \left(\frac{N_{\mbox{\scriptsize Se}}}{N_{\mbox{\scriptsize Se}}+N_{\mbox{\scriptsize Pt}}}\right) \,& \mu_{\mbox{\scriptsize Se}}^{\mbox{\tiny Se-rich}} + \nonumber \\ 
& \left(\frac{N_{\mbox{\scriptsize Pt}}}{N_{\mbox{\scriptsize Se}}+N_{\mbox{\scriptsize Pt}}}\right) \, \mu_{\mbox{\scriptsize Se}}^{\mbox{\tiny Pt-rich}} \label{muSeGeneral} \\[8mm]
\mu_{\mbox{\scriptsize Pt}} = \left(\frac{N_{\mbox{\scriptsize Se}}}{N_{\mbox{\scriptsize Se}}+N_{\mbox{\scriptsize Pt}}}\right) \,& \mu_{\mbox{\scriptsize Pt}}^{\mbox{\tiny Se-rich}} + \nonumber \\ 
& \left(\frac{N_{\mbox{\scriptsize Pt}}}{N_{\mbox{\scriptsize Se}}+N_{\mbox{\scriptsize Pt}}}\right) \, \mu_{\mbox{\scriptsize Pt}}^{\mbox{\tiny Pt-rich}} \label{muPtGeneral}
\end{align}
where $N_{\mbox{\scriptsize Se}}$ and $N_{\mbox{\scriptsize Pt}}$ are the total number of Se and Pt atoms in the film, respectively, and values for the Se-rich and Pt-rich $\mu_{\mbox{\scriptsize i}}$'s are taken from Table \ref{RichChemPotentials}.

\subsubsection{Energies of Interacting Defects \label{InteractingDefectsSubSection}}

Table~\ref{ConfigurationEnergies} lists the geometric parameters employed in the DFT calculations and the DFT energies obtained for the six isolated defects and the twenty configurations of interacting defects observed with scanning tunneling microscopy. The formation and group energies in Table ~\ref{ConfigurationEnergies} are calculated at two conditions for film growth: Se-rich, and Pt-rich. For a film of arbitrary composition between these conditions, the formation or group energy ($D_d$ or $V_m$) in Equations~\ref{PottsModelEfj} and \ref{PottsModelEgj} is calculated using linear interpolation:
	\begin{align}
    D_{\mbox{\scriptsize d}} = \left(\frac{N_{\mbox{\scriptsize Se}}}{N_{\mbox{\scriptsize Se}}+N_{\mbox{\scriptsize Pt}}}\right) \,& D_{\mbox{\scriptsize d}}^{\mbox{\tiny Se-rich}} + \nonumber \\ 
		& \left(\frac{N_{\mbox{\scriptsize Pt}}}{N_{\mbox{\scriptsize Se}}+N_{\mbox{\scriptsize Pt}}}\right) \, D_{\mbox{\scriptsize d}}^{\mbox{\tiny Pt-rich}} \label{DdGeneral} \\[8mm]
	V_{\mbox{\scriptsize m}} = \left(\frac{N_{\mbox{\scriptsize Se}}}{N_{\mbox{\scriptsize Se}}+N_{\mbox{\scriptsize Pt}}}\right) \,& V_{\mbox{\scriptsize m}}^{\mbox{\tiny Se-rich}} + \nonumber \\ 
		& \left(\frac{N_{\mbox{\scriptsize Pt}}}{N_{\mbox{\scriptsize Se}}+N_{\mbox{\scriptsize Pt}}}\right) \, V_{\mbox{\scriptsize m}}^{\mbox{\tiny Pt-rich}} \label{VmGeneral}
	\end{align}
	The values for the Se-rich and Pt-rich $D_{\mbox{\scriptsize d}}$'s and $V_{\mbox{\scriptsize m}}$'s in these expressions are provided in Table \ref{ConfigurationEnergies}.
	
	In Equations~\ref{muSeGeneral}~--~\ref{VmGeneral}, the quantities $N_{\mbox{\scriptsize Se}}$ and $N_{\mbox{\scriptsize Pt}}$ represent the total numbers of Se and Pt atoms, respectively, in the film. They are calculated from the numbers of Se and Pt atoms in a stoichiometric film and the changes in these numbers caused by introducing defects:
	\begin{align}
		N_{\mbox{\scriptsize Pt}} = 2N - \left( n_{\mbox{\scriptsize B}} + n_{\mbox{\scriptsize D}} + n_{\mbox{\scriptsize E}} + n_{\mbox{\scriptsize F}} \right) \\[8mm]
		N_{\mbox{\scriptsize Se}} = 4N - \left( n_{\mbox{\scriptsize A}} + n_{\mbox{\scriptsize C}} - n_{\mbox{\scriptsize E}} + n_{\mbox{\scriptsize F}} \right)
	\end{align}
	Here, $N$ is the number of lattice sites used in the Monte Carlo simulation and the $n_i$'s are the numbers of each defect type. Each lattice site represents a unit cell of Pt$_2$Se$_4$; this represents a monolayer of 2D PtSe$_2$ which is made up of two layers of the PtSe$_2$ formula unit.

\begin{table*}[h!]
\footnotesize
\begin{tabularx}{0.95\textwidth} { 
  | >{\raggedright\arraybackslash}c 
  | >{\raggedright\arraybackslash}c
  | >{\raggedright\arraybackslash}c 
  | >{\raggedright\arraybackslash}c
  | >{\raggedright\arraybackslash}c 
  | >{\raggedright\arraybackslash}c 
  | >{\raggedright\arraybackslash}c
  | >{\raggedright\arraybackslash}c |
  } \hline 
   \multicolumn{1}{|c|}{ Defect } & \multicolumn{1}{c|}{Defect type} & \multicolumn{1}{c|}{Supercell} & \multicolumn{1}{c|}{K-mesh} & \multicolumn{2}{c}{Formation energy} & \multicolumn{2}{|c|}{Group energy} \\ 
\multicolumn{1}{|c|}{identifier} & \multicolumn{1}{c|}{(point defect} & \multicolumn{1}{c|}{size} & & \multicolumn{2}{c}{at synthesis} & \multicolumn{2}{|c|}{(formation+interaction) at} \\
  & \multicolumn{1}{c|}{or combination)} & & & \multicolumn{2}{c}{composition (eV)} & \multicolumn{2}{|c|}{synthesis composition (eV) } \\
  \cline{5-8}  \rule[-.3\baselineskip]{0mm}{2mm}
  & & & & \multicolumn{1}{X|}{\hspace{3mm} $D_{\mbox{\scriptsize d}}^{\mbox{\tiny Se-rich}}$ } & \multicolumn{1}{X|}{\hspace{3mm} $D_{\mbox{\scriptsize d}}^{\mbox{\tiny Pt-rich}}$ } &  \multicolumn{1}{X|}{\hspace{3mm} $V_{\mbox{\scriptsize m}}^{\mbox{\tiny Se-rich}}$ } & \multicolumn{1}{X|}{\hspace{3mm} $V_{\mbox{\scriptsize m}}^{\mbox{\tiny Pt-rich}}$ }  \\[1mm]  \hline
A & $V_{\mbox{\scriptsize Se1a}}$ & $5 \times 5$ & $5 \times 5 \times 1$ & 1.464 & 0.877 & & \\
B & $V_{\mbox{\scriptsize Pt1}}$ & $5 \times 5$ & $5 \times 5 \times 1$ & 1.353 & 2.508 & & \\
C & $V_{\mbox{\scriptsize Se1b}}$ & $5 \times 5$ & $5 \times 5 \times 1$ & 1.884 & 1.307 & & \\
D & $V_{\mbox{\scriptsize Pt2}}$ & $5 \times 5$ & $5 \times 5 \times 1$ & 1.353 & 2.508 & & \\
E & $\mbox{Se}_{\mbox{\scriptsize Pt1}}$ & $5 \times 5$ & $5 \times 5 \times 1$ & $-0.064$ & 1.668 & & \\
F & $V_{\mbox{\scriptsize Se1b}}V_{\mbox{\scriptsize Pt1}}$ & $5 \times 5$ & $5 \times 5 \times 1$ & 2.811 & 3.227 & & \\
1 & $2V_{\mbox{\scriptsize Se1a}}$ & $5 \times 5$ & $5 \times 5 \times 1$ & & & 2.4865 & 1.5674 \\
2 & $3V_{\mbox{\scriptsize Se1a}}$ & $5 \times 5$ & $5 \times 5 \times 1$ & & & 3.987 & 2.231 \\
3 & $2V_{\mbox{\scriptsize Pt1}}$ & $5 \times 5$ & $5 \times 5 \times 1$ & & & 2.387 & 4.869 \\
4 & $2V_{\mbox{\scriptsize Pt1}}$ & $5 \times 5$ & $5 \times 5 \times 1$ & & & 2.531 & 5.278 \\
5 & $2V_{\mbox{\scriptsize Pt1}}$ & $5 \times 5$ & $5 \times 5 \times 1$ & & & 2.598 & 5.076 \\
6 & $2V_{\mbox{\scriptsize Se1b}}$ & $5 \times 5$ & $5 \times 5 \times 1$ & & & $-0.155$ & 3.268 \\
7 & $2V_{\mbox{\scriptsize Se1b}}$ & $7 \times 7$ & $3 \times 3 \times 1$ & & & 3.387 & 2.693 \\
8 & $2V_{\mbox{\scriptsize Pt2}}$ & $9 \times 9$ & $3 \times 3 \times 1$ & & & 3.381 & 2.653 \\
9 & $2V_{\mbox{\scriptsize Pt2}}$ & $9 \times 9$ & $3 \times 3 \times 1$ & & & 2.981 & 3.767 \\
10 & $2V_{\mbox{\scriptsize Pt2}}$ & $11 \times 11$ & $3 \times 3 \times 1$ & & & 2.811 & 3.227 \\
11 & $2V_{\mbox{\scriptsize Pt2}}$ & $11 \times 11$ & $3 \times 3 \times 1$ & & & 2.563 & 5.291 \\
12 & $2\mbox{Se}_{\mbox{\scriptsize Pt1}}$ & $5 \times 5$ & $5 \times 5 \times 1$ & & & 2.570 & 5.297 \\
13 & $2\mbox{Se}_{\mbox{\scriptsize Pt1}}$ & $5 \times 5$ & $5 \times 5 \times 1$ & & & 2.671 & 5.115 \\
14 & $2\mbox{Se}_{\mbox{\scriptsize Pt1}}$ & $5 \times 5$ & $5 \times 5 \times 1$ & & & 2.708 & 5.146 \\
15 & $5\mbox{Se}_{\mbox{\scriptsize Pt1}}$ & $11 \times 11$ & $3 \times 3 \times 1$ & & & $-0.133$ & 3.391 \\
16 & $V_{\mbox{\scriptsize Se1b}}V_{\mbox{\scriptsize Pt1}}$ & $5 \times 5$ & $5 \times 5 \times 1$ & & & $-0.139$ & 3.178 \\
17 & $V_{\mbox{\scriptsize Se1b}}V_{\mbox{\scriptsize Pt1}}$ & $5 \times 5$ & $5 \times 5 \times 1$ & & & 1.327 & 2.637 \\
18 & $V_{\mbox{\scriptsize Se1b}}V_{\mbox{\scriptsize Pt1}}$ & $5 \times 5$ & $5 \times 5 \times 1$ & & & $-0.368$ & 9.802 \\
19 & $\mbox{Se}_{\mbox{\scriptsize Pt1}}V_{\mbox{\scriptsize Se1a}}$ & $5 \times 5$ & $5 \times 5 \times 1$ & & & 2.835 & 3.831 \\
20 & $V_{\mbox{\scriptsize Pt1}}V_{\mbox{\scriptsize Pt2}}$ & $7 \times 7$ & $3 \times 3 \times 1$ & & & 2.715 & 5.103 \\ \hline
\end{tabularx}
\normalsize
\caption{Defects types, their corresponding supercell and K-mesh sizes for the DFT calculations, and their formation or group energies at two synthesis conditions. }
\label{ConfigurationEnergies} 
\end{table*}

\end{appendices}

\clearpage


\bibliographystyle{apsrev4-2}
\bibliography{PtSe2ref} 

\end{document}